\documentclass[11pt]{article}
\usepackage[dvipsnames]{xcolor} 
\usepackage[hidelinks]{hyperref}
\usepackage{enumitem}
\usepackage{amsmath,amssymb,epsf,cite,graphicx,subfigure}

\usepackage{amsmath,braket,tikzsymbols}
\usepackage[bbgreekl]{mathbbol}
\usepackage{bbold}
\usepackage{subfigure}

\numberwithin{equation}{section}

\definecolor{airforceblue}{rgb}{0.36, 0.54, 0.66}
\newcommand{\beq}{\begin{equation}}
\newcommand{\eeq}{\end{equation}}

\usepackage[mathscr]{euscript}

\begin{document}
\baselineskip=15.5pt
\pagestyle{plain}
\setcounter{page}{1}

\begin{center}
{\LARGE \bf Comparing top-down and bottom-up holographic defects and boundaries}
\vskip 1cm

\textbf{William Harvey$^{a,1}$, Kristan Jensen$^{b,1}$, and Takahiro Uzu$^{c,1,2}$}

\vspace{0.5cm}

{\small $^1$Department of Physics and Astronomy, University of Victoria, Victoria, BC V8W 3P6, Canada\\}

{\small $^2$Faculty of Education, University of Cambridge, 184 Hills Road, Cambridge, CB2 8PQ, United Kingdom}

\vspace{0.3cm}

{\tt \small ${}^a$wharvey@uvic.ca,}
{\tt  \small ${}^b$kristanj@uvic.ca }
{\tt \small ${}^c$tu234@cam.ac.uk \\}

\medskip

\end{center}

\vskip1cm

\begin{center}
{\bf Abstract}
\end{center}

In this work we consider domain walls and end-of-the-world branes in AdS/CFT, holographically dual to codimension-one conformal defects and conformal boundaries respectively. In this setting there is an analogue of the ``bulk point'' singularity in boundary correlation functions, which we use to compare top-down and bottom-up constructions of these systems. For example, for a range of parameters the D3/D5 boundary CFT cannot be imitated by a tensionful end-of-the-world brane coupled to Einstein gravity, and in another range it can be modeled with a negative tension brane. Along the way we compute the central charge $b$ for the M2/M5 boundary CFT.

\hspace{.3cm}

\newpage

\tableofcontents

\section{Introduction}

The best understood version of holographic duality is the AdS/CFT correspondence~\cite{Maldacena:1997re}, which states that consistent theories of quantum gravity in asymptotically anti-de Sitter (AdS) spacetimes are dual to conformal field theories (CFTs) that, roughly speaking, live on the AdS boundary. These dualities equate the strong coupling limit of a CFT with the semiclassical limit of a dual gravity description. As such AdS/CFT geometrizes otherwise intractable and inaccessible phenomena in strongly correlated field theories. 

In this work we study holographic dualities where the CFT has been modified by the introduction of a conformal defect or boundary. In the bulk, a defect is dual to a domain wall separating two distinct regions of the bulk spacetime, while a boundary is dual to an ``end of the world brane'' (ETW brane) on which the bulk spacetime ends. In string theoretic realizations of holography, what we call ``top-down'' in this manuscript, the bulk is 10- or 11-dimensional string/M-theory and these domain walls and/or ETW branes correspond to dynamical objects, like branes or non-trivial configurations of the supergravity fields. See e.g.~\cite{Bak:2003jk,DHoker:2006vfr,Gomis:2006cu,DHoker:2007zhm,DHoker:2007hhe,DHoker:2009lky,Kim:2009wv,Bachas:2013vza,Bobev:2013yra}. These domain walls and boundaries are often intricate; in many cases they preserve a large amount of supersymmetry and as such offer an opportunity for precision tests on holographic duality in this arena. In particular, string theory realizations of domain walls and boundaries often involve the compact directions of the bulk spacetime in a crucial way.

An example to keep in mind is the holographic dual of the D3/D5 boundary CFT, in which $N_3$ D3 branes end on $N_5$ D5 branes in such a way as to preserve half of the superconformal symmetry of the $U(N_3)$ super-Yang Mills theory living on the D3 branes. The D5 branes correspond to a certain conformal boundary condition uncovered by Gaiotto and Witten~\cite{Gaiotto:2008sa} (see also~\cite{DeWolfe:2001pq}). For a large number of D5 branes the bulk undergoes a geometric transition and the dual geometry is locally AdS$_4\times \mathbb{S}^2\times\mathbb{S}^2\times \Sigma$ with $\Sigma$ a strip and is determined in terms of two harmonic functions on the strip~\cite{DHoker:2007hhe}; the 10d geometry ends smoothly on one end of the strip, and becomes AdS$_5\times\mathbb{S}^5$ at the other. An impressive precision check of the duality has been performed in this example, namely the computation of the ``boundary $F$''~\cite{Gaiotto:2014gha} in both 10d type IIB supergravity~\cite{Estes:2014hka,Raamsdonk:2020tin} and in the dual boundary CFT~\cite{Raamsdonk:2020tin}, and of course the two computations agree on the nose.

Because top-down realizations of defects and boundaries are intricate and complex, much recent attention has been devoted to simpler ``bottom-up'' models of defects and boundaries. These models were introduced with the aspiration of describing coarse features of proper string theoretic realizations of interfaces and boundaries without the complications of the compact directions, supergravity fluxes, and supersymmetry. Indeed the basic bottom-up models on the market involve a tensionful brane -- either a codimension-one domain wall~\cite{Karch:2000ct}, or a boundary~\cite{Takayanagi:2011zk} -- coupled to Einstein gravity with negative cosmological constant. The bulk geometry is then Einstein away from the domain wall or boundary, and typically these research efforts do not include compact directions. These models have been en vogue of late for various problems in AdS/CFT, including boundary CFT realizations of black hole microstates and the black hole information paradox.

There is a history of using bottom-up models of bulk gravity particularly in the context of ``applied AdS/CFT,'' where instead of considering the effective supergravity descriptions of AdS$_{d+1}\times X$ in string/M-theory, with their attendant AdS field content and interactions, one considers bulk effective field theories of Einstein gravity coupled to matter with simple interactions. In some cases these models can even be embedded into string theory or upgraded to systematic approximations of top-down constructions. By comparison bottom-up models of interfaces and ETW branes appear to be unsystematic and ad hoc approximations to top-down models, although see~\cite{Harvey:2023pdv,Sugimoto:2023oul}. Perhaps a suitable coarse-graining of top-down constructions leads to the bottom-up ones, but at the present time such a simplifying map does not exist.

This raises a question, namely, how well do bottom-up models match the physics of their top-down counterparts? The basic goal of this manuscript is to begin answering that question. The way we do this is through a quantity we call the ``light crossing time'' which we notate as $\phi_b$. This quantity characterizes the causal structure of the holographic dual to a defect or boundary CFT. It was originally observed in~\cite{Reeves:2021sab} and there was called the ``causal depth.'' We will momentarily describe what this parameter is and why it is interesting, but for now the important point is threefold: (1) this quantity can be defined in the dual CFT, (2) bottom-up models can be described equivalently in terms of their tension or this light crossing time, and (3) this quantity is computable in top-down examples in terms of dual field theory parameters. As such, we can use it to match a bottom-up model to a top-down construction, or to show that no such matching exists. (See also~\cite{Bachas:2024nvh,Bachas:2025brp} for other discussions of CFT notions of tension of bulk objects.)

Having discussed its usefulness, let us back up and explain what the light crossing time is. In AdS/CFT, it is sometimes the case that points on the boundary can be connected by bulk null geodesics. For instance these trajectories exist when two points on the boundary are null separated, leading to the usual Regge singularity of CFT correlation functions. By either a geometric optics approximation, or by rewriting boundary-to-boundary propagators as a sum over worldlines, we see that these geodesics imply exact or approximate singularities to correlation functions of the dual CFT. There are also situations where these trajectories exist at cross-ratios that do not correspond to an expected singularity of a CFT correlator. In global AdS spacetime the prototypical example of this idea is the ``looking for a bulk point'' singularity of~\cite{Maldacena:2015iua}, wherein four points on the AdS boundary are connected by null geodesics to a bulk point. In the supergravity approximation the dual CFT four-point function has a singularity at the cross-ratios corresponding to that configuration. Those cross-ratios do not correspond to any expected singularity in CFT, and indeed it has been argued that this singularity is only approximate and is smoothed out at the string scale. From the point of view of the conformal bootstrap program these singularities are useful as they sharply characterize the emergence of bulk locality in terms of CFT data.

There is a defect or boundary CFT analogue of the ``looking for a bulk point'' singularity as noted in~\cite{Reeves:2021sab}. This singularity, also expected to be only approximate, arises from null geodesics that connect two points on the AdS boundary after either bouncing off an end-of-the-world brane (if we are considering the holographic dual of a boundary CFT) or passing through a bulk domain wall (if we are considering the dual of a defect CFT). Two-point functions of a boundary or defect CFT are characterized by a single cross-ratio, and this singularity appears at cross-ratios other than those expected from the OPE or an analogue of the Regge limit. 

The cross-ratio corresponding to this geodesic can be interpreted as a time in the following way. Suppose that we place the boundary CFT on the hemisphere, or if we are considering a codimension-one defect, on the sphere with the defect extended along the equator. In the boundary case, after placing the two operator insertions at the pole of the hemisphere, the ``light crossing time'' is half the boundary time it takes for the null geodesic to pass from the pole back to itself through the bulk. As we will see, this light crossing time is generically, though not always finite, and it characterizes a sense in which holographic boundary or defect CFTs have a local bulk dual. 

We compute this light crossing time in the simple bottom-up models of gravity coupled to a tensionful brane (the Karch-Randall model~\cite{Karch:2000ct}), or ending on a tensionful brane (Takayanagi's model of holographic BCFT~\cite{Takayanagi:2011zk}), as well as in a variety of top-down constructions like the D3/D5 system and M2/M5 boundary CFT. The top-down calculations are new results. For codimension-one defects, we find that the bulk domain walls fall into one of two classes: either they can be imitated by a Karch-Randall brane of positive tension, or they cannot be imitated by any such brane. For boundaries there is a slightly richer story, whereby top-down models can be modeled by an end-of-the-world brane with negative tension in some range of boundary parameters, positive tension in another, and no such tensionful end-of-the-world brane in the remainder.

From here we further explore those top-down constructions whose light crossing times can be matched to that of a bottom-up description. Conformal defects or boundaries can be characterized by a ``defect entropy'' or ``boundary entropy,'' like $g$ of 1+1-dimensional boundary CFT~\cite{Affleck:1991tk} and its analogues in higher dimensions~\cite{Nozaki:2012qd,Estes:2014hka,Gaiotto:2014gha,Jensen:2015swa,Kobayashi:2018lil}. In some scenarios this entropy has been shown to be a measure of the number of degrees of freedom associated with a defect or boundary, insofar as it decreases from UV to IR under RG flows triggered by operators on the boundary or defect when the boundary or defect is 0+1~\cite{Friedan:2003yc} or 1+1-dimensional~\cite{Jensen:2015swa} (see also~\cite{Casini:2018nym}). In bottom-up models of defects and boundaries the defect/boundary entropy has a simple qualitative behavior as a function of tension: it increases monotonically with tension, vanishing for a tensionless Karch-Randall or end-of-the world brane~\cite{Azeyanagi:2007qj,Kobayashi:2018lil}. We compare this qualitative behavior with that of top-down constructions, relying on previous results for the defect/boundary entropy where we can and computing it in examples where it was unknown. 

For top-down defects we find that defect entropy behaves in a way that resembles the Karch-Randall model, behaving monotonically with the effective tension (as matched through the light crossing time) and with the same sign, while vanishing at zero effective tension. See Figs.~\ref{d4DCFTCompFig} and~\ref{d2DCFTCompFig}. For top-down boundaries we also find that boundary entropy also grows monotonically with the effective tension, agreeing quite well with bottom-up results in the regime of light crossing time that, in the bottom-up context, would correspond to a negative tension ETW brane. See Figs.~\ref{d4BCFTCompFig} and~\ref{d3BCFTCompFig}.

We collect results for a modest inventory of top-down defects and boundaries. For defects we consider Janus interfaces of various types, as well as the D3/D5 defect CFT. As for top-down duals to boundary CFT, we study the D3/D5 and M2/M5 intersections. The boundary entropy for the latter, its $b$-type central charge, was previously unknown and we compute it in this work.

The remainder of this manuscript is organized as follows. In Section~\ref{S:lightCrossing} we discuss aspects of the light crossing time, both boundary and bulk points of view. We also review the gravity backgrounds and light crossing time for the simple Karch-Randall and Takayanagi models of holographic defect and boundary CFT. We compute the light crossing time for a variety of holographic defect CFTs in Section~\ref{Interfaces} and boundary CFTs in Section~\ref{Boundaries}, compare qualitative features of the entropy as we go. We wrap up with a Discussion in Section~\ref{discussion}, and relegate some details concerning perturbative Janus interfaces and top-down defect/boundary entropy to the Appendix.

\section{Light crossing time}
\label{S:lightCrossing}

In this Section we explain the physics of the light crossing time $\phi_b$ from both sides of an AdS/DCFT or AdS/BCFT correspondence. We start on the gravity side, where we show that for a holographic BCFT $\phi_b$ refers to a time of travel for null geodesics to go from the AdS boundary to the end of the bulk spacetime. Then we show that the same quantity controls the location of an approximate singularity in the correlation functions of a holographic DCFT or BCFT. In both cases, $\phi_{b}$ is a defining characteristic of holographic DCFTs and BCFTs no matter their underlying details or bulk matter content. 

\subsection{From geometry}
\label{geometric}

Consider the gravity dual of a flat space $d$-dimensional DCFT or BCFT with a flat defect or boundary. The DCFT or BCFT lives in the spacetime $ds^2 = -dt^2 + dx_{\perp}^2 + d\vec{x}^2$ with the defect or boundary located at $x_{\perp} = 0$, preserving an $SO(d-1,2)$ subgroup of the full $SO(d,2)$ conformal symmetry. For now we will take the dual to be $d+1$-dimensional. The residual conformal symmetry then constrains the bulk metric to take the form
\begin{equation}
    \label{MainMetric}
    ds^{2} = d \rho^{2} + e^{2 A(\rho)} \left(\frac{-dt^{2} + dz^{2} + d \mathbf{x}^{2}}{z^{2}} \right)\,,
\end{equation}
Here $\rho$ is a radial coordinate which we take to have extent $\rho \in (\rho_0,\infty)$ and the warpfactor $A(\rho)$ is determined by the matter profile through the Einstein's equations.  For the dual to a DCFT we have $\rho\in \mathbb{R}$ and the geometry has a conformal boundary given by the union of three parts. The first two are equivalent to half of $d$-dimensional Minkowski space, reached as $\rho \to \pm \infty$ near which $A(\rho)$ grows linearly with $|\rho|$, so that the geometry is asymptotically AdS$_{d+1}$ in those regions. The third is a codimension-two surface equivalent to $d-1$-dimensional Minkowski space that comes from taking $z\to 0$. In the dual DCFT this last region is the conformal defect, while the half-spaces reached as $\rho\to \pm\infty$ are the two halves of Minkowski space on either side. On the conformal boundary the radial coordinate $z$ of the AdS$_d$ slices parameterizes the direction $x_{\perp}$ perpendicular to the defect. For the dual of a BCFT there are only two components of the conformal boundary reached as $\rho\to\infty$ and $z\to 0$. The latter corresponds to the boundary of the BCFT and the former to the half of Minkowski space ending on it. 

We wish to study null geodesics emanating from the conformal boundary in spacetimes of this form. We can do this by a simple argument from~\cite{Harvey:2023pdv} alluded to in~\cite{Reeves:2021sab}. We reparameterize the radial coordinate as $d\phi = -\frac{d\rho}{e^{A(\rho)}}$ and take $\rho\to\infty$ to correspond to $\phi=0$. Then the bulk spacetime~\eqref{MainMetric} is conformal to a wedge of $d+1$-dimensional Minkowski space,
\begin{equation}
  ds^2=\frac{e^{2A(\rho)}}{z^2}\left(-dt^2+d\mathbf{x}^2+dz^2+z^2d\phi^2 \right)\,,
\end{equation}
and null geodesics are straight lines in these coordinates. The angular extent of the wedge is
\begin{equation}\label{PhiB}
    \phi_b = \int_{\rho_0}^{\infty} d\rho\,e^{-A(\rho)}\,.
\end{equation}
For a DCFT the surfaces $\phi = 0$ and $\phi=\phi_b$ correspond to the two halves of flat space adjoining the defect, while for a BCFT the bulk spacetime ends on surface $\phi = \phi_b$. The defect or boundary is located at the origin of the wedge at $z=0$.

Null geodesics emanating from the boundary then fall into one of two types. Either they start at the defect or boundary and shoot into the bulk, or they begin away from the defect or boundary, say on the $\phi=0$ surface. If $\phi_b <\pi$ these geodesics can traverse the bulk to the $\phi=\phi_b$ surface in finite boundary time. 

In fact there is a similar story when $\phi_b\geq \pi$. To see this we consider the gravity dual of a DCFT or BCFT on either a sphere or hemisphere respectively, with the defect/boundary wrapping the equator. The bulk geometry is now the same as above but with global AdS$_d$ slices, which we can parameterize as
\begin{equation}
    ds^2 = d\rho^2 + e^{2A(\rho)}ds^2_{\text{ global}}\,, \qquad ds^2_{\text{global}} = \frac{-d\tau^2+d\theta^2+\sin^2(\theta) \,d\Omega_{d-2}^2}{\cos^2(\theta)}\,.
\end{equation}
Here $\theta \in [0,\pi/2]$ so that the slices are conformal to global time $\tau$ times a hemisphere. This geometry is still a wedge of flat space, 
\begin{equation}
    ds^2 = \frac{e^{2A(\rho)}}{\cos^2(\theta)}\left( -d\tau^2 + d\theta^2 + \sin^2(\theta) d\Omega_{d-2}^2 + \cos^2(\theta) d\phi^2\right)\,.
\end{equation}
Now consider a null geodesic emanating from a point on the $\phi=0$ surface going into the bulk. By the residual conformal invariance we can take this point to be at $\tau=0$ and at the pole of the hemisphere , $\theta = 0$, and for the ensuing motion to take place in $(\tau,\phi)$. The resulting path takes a time $\Delta \tau = \phi_b$ to travel from this point to the $\phi=\phi_b$ surface. 

In this way we can think of $\phi_b$ as a light-crossing time from the part of the conformal boundary at $\phi = 0$ to the other surface at $\phi = \phi_b$, whether it is another half of the conformal boundary as for a DCFT or an end-of-the-world brane as for a BCFT. From this point of view the time of travel from the conformal boundary to itself is $\phi_b$ for a DCFT, and $2\phi_b$ for a BCFT. This result also explains a feature of our Poincar\'e sliced analysis above: since a global time of $\pi$ corresponds to infinite Poincar\'e time, if $\phi_b<\pi$ then this trajectory takes a finite amount of Poincar\'e time to arrive at the $\phi=\phi_b$ surface, whereas if $\phi_b\geq \pi$  then this geodesic is not contained within the Poincar\'e sliced patch.

If we consider the two-point function of a local CFT operator, with both insertions located at the pole $\theta = 0$, then by writing this two-point function as a sum over bulk worldlines we expect this correlation function to have a singularity when the two insertions are separated by twice this light-crossing time $2\phi_b$. In the next Subsection we will see that this is precisely the case.

\subsection{From correlation functions}

There are two slightly different approaches one can use to compute correlation functions for a BCFT or DCFT. The first is via the OPE, and to sum over bulk primary operators. The second is to relate the bulk operators to the boundary ones, and to instead sum over the latter via the boundary operator expansion. These equivalent approaches are concisely stated below:
\begin{equation}\label{channels}
    \sum \limits_{i} a_{i} g^{B}_{\Delta_{i}} (\xi) = \sum \limits_{I} b_{I} g^{b}_{\hat{\Delta}_{I}} (\xi), 
\end{equation}
where $a_{i} = A_{i} C_{i}$ is the product of the bulk OPE coefficient and the one-point function coefficient for the operator $\mathcal{O}_{i}$, and $b_{I} = B_{I}^{2}$ is the square of the BOE coefficient of the boundary operator $\hat{\mathcal{O}}_{I}$. The $g_{\Delta_{i}}^{B}$ and $g_{\hat{\Delta}_{I}}^{b}$ are the bulk and boundary conformal blocks, respectively; the cross-ratio $\xi$ is defined as
\begin{equation}
    \xi = \frac{(x - y)^{2}}{4x_{\perp}y_{\perp}} \,.
\end{equation}
These scalar conformal blocks are obtained by solving the Casimir equation for the full and reduced conformal symmetries, respectively
\begin{equation}\label{blocks}
\begin{split}
    g^{B}_{\Delta_i} (\xi) &= \xi^{\Delta_i/2 - \Delta} {}_2 F_{1} \left(\frac{\Delta_i}{2}, \frac{\Delta_i}{2}; \Delta_i - \frac{d}{2} + 1; - \xi \right) \,,
    \\
    g^{b}_{\hat{\Delta}_I}(\xi) &= \xi^{-\hat{\Delta}_I} {}_2F_{1} \left(\hat{\Delta}_I, \hat{\Delta}_I - \frac{d}{2}+1; 2 \hat{\Delta}_I + 2 - d; -\xi^{-1} \right)\,.
\end{split}
\end{equation}
See~\cite{McAvity:1995zd} for the full details.

Given a holographic model of the vacuum state of a BCFT dual to a line element of the form~\eqref{MainMetric} for $\rho \geq \rho_0$, and a minimally coupled scalar field $\varphi$ dual to a scalar operator $\mathcal{O}$ of dimension $\Delta$, we now compute the spectrum of defect/boundary operator dimensions $\hat{\Delta}_{I}$ in the spectrum of $\mathcal{O}$ as well as the asymptotics of the BOE coefficients. (The authors of~\cite{Reeves:2021sab} did so for a cutout of AdS.) In this way we show that holographic BCFTs have an approximate singularity in correlation functions controlled by the same quantity $\phi_b$ that appeared above. To do so, we introduce the radial coordinate $r$ such that the cross ratio becomes $\xi=\frac{(1-r)^2}{4 r}$; this maps $\xi \in (0, \infty)$ with $r \in (0,1)$. The boundary block $g^{b}$ is therefore rewritten as
\begin{equation}
 g^{b}_{\hat{\Delta}} = (4 r)^{\hat{\Delta}} {}_2 F_{1} \left(\hat{\Delta}, \frac{d-1}{2}; \hat{\Delta} - \frac{d}{2} + \frac{3}{2}; r^{2} \right).
\end{equation}
Note that there are singularities at: $r = 0, r \pm 1, $ and $r = \infty$. If there is to be any other emergent approximate singularity, then this must occur upon summing the blocks, with the most significant contributions coming from the heaviest operators. 

In the large dimension limit, the boundary block simplifies to
\begin{equation}\label{bdyblock}
    \lim_{\hat{\Delta}\rightarrow \infty} g^{b}_{\hat{\Delta}} (r) = \frac{(4 r)^{\hat{\Delta}}}{(1 - r^{2})^{(d-1)/2}}.
\end{equation}
The minimally coupled scalar field $\varphi$ obeys
\begin{equation}
    \Box\varphi-m^2\varphi = 0 \,.
\end{equation}
with $m^2 = \Delta(\Delta-d)$. To extract the $\hat{\Delta}_I$ and the BOE coefficients we find the most general solution to this equation~\cite{Aharony:2003qf}, expanding $\varphi$ into AdS harmonics $\Psi_n(x)$ obeying $\Box_d\Psi_n =m_n^2 \Psi_n$, as $\varphi=\sum_n c_{n} \Psi_n(x)\psi_n(\rho)$ for $n \in \mathbb{Z}_{+}$. Each $\psi_{n}$ is dual to a boundary operator $\hat{\mathcal{O}}_n$ with $m_n^2 = \hat{\Delta}_n(\hat{\Delta}_n-(d-1))$. We proceed to solve the Klein-Gordon equation in the limit that the boundary dimensions $\hat{\Delta}_n$ are large, so that
\begin{equation}
   \psi''_{n} + d A' \psi'_{n} - m^2\psi_n   + e^{-2A} \hat{\Delta}_{n}^{2}\psi_n =0 \,.
\end{equation}
From here, we perform a WKB ansatz
\begin{equation}
    \psi_n(\rho)=\exp{\left(\frac{1}{\epsilon} \sum_{m=0}^\infty \epsilon^m S_m(\rho) \right)} \,,
\end{equation}
where we perform the expansion in $\epsilon=\frac{1}{\hat{\Delta}_n}\ll 1$. Substituting this into the Klein-Gordon equation
\begin{equation}
    \left(\sum_{m=0}^\infty \epsilon^m S'_m(\rho) \right)^2+\epsilon\left(\sum_{m=0}^\infty \epsilon^m S''_m(\rho) \right)+dA'\epsilon\left(\sum_{m=0}^\infty \epsilon^m S'_m(\rho) \right)-m^2\epsilon^2+e^{-2A}\hat{\Delta}_n^2=0 \,
\end{equation}
we can now solve this order-by-order. To leading-order, we find
\begin{equation}
    S_0(\rho)=\pm i \hat{\Delta}_n\int_{\rho}^{\infty}d\rho' \,e^{-A(\rho')}  \,.
\end{equation}
To first order, using $S_{0}$, we find
\begin{equation}
    S_1(\rho)=\frac{1-d}{2}A+(\text{constant}) \,,
\end{equation}
and this is all we need. The WKB approximation then gives two solutions to the radial problem
\begin{equation}
    \psi_n(\rho) = \exp{\left(\pm i\hat{\Delta}_n\int_{\rho}^{\infty} d\rho' \,e^{-A(\rho')}+\frac{1-d}{2}A\right)} \,.
\end{equation}
The boundary condition on $\varphi$ will depend on the details of the ETW brane on which spacetime ends at $\rho=\rho_0$. We remain agnostic as to this boundary condition, so that $\psi_n$ is some real linear combination of the two solutions
\begin{equation}
\label{E:WKBsolution}
    \psi_n(\rho)\propto e^{\frac{1-d}{2}A}\cos{\left(\hat{\Delta}_n \int_{\rho}^{\infty}d\rho' e^{-A(\rho')}+\delta_n\right)} \,,
\end{equation}
for some phase shift $\delta_n$. $\psi_n$ decays near the asymptotically AdS boundary, while at the ETW 
\begin{equation}
    \psi_n(\rho_0) \propto e^{\frac{1-d}{2}A}\cos\left(\hat{\Delta}_n\phi_b+\delta_n\right)\,.
\end{equation}
Under a Dirichlet condition we infer that $\hat{\Delta}_n$ is regularly spaced with
\begin{equation}
\label{E:spacing}
    \hat{\Delta}_{n+1} = \hat{\Delta}_n + \frac{\pi}{\phi_b}\,,
\end{equation}
and the same holds for any linear condition on $\psi_n$ and $\psi_n'$.

To obtain the BOE coefficients, one starts from the bulk two-point function $\langle \varphi (x) \varphi (y) \rangle $, and again expands into AdS harmonics. Following \cite{Penedones:2016voo} and using the AdS$_{d}$ bulk-to-bulk propagator 
\begin{equation}\label{b2bp}
    G_{\Delta}^{AdS_{d}} (\xi) = C_{\Delta, d-1} 2^{-2 \Delta} \xi^{-\Delta} {}_2 F_{1} (\Delta, \Delta - d/2 + 1, 2 \Delta - d + 2, -1/\xi),
\end{equation}
with
\begin{equation}
    C_{\Delta_{n},d-1} = \frac{\Gamma (\Delta)}{2 \pi^{\frac{d-1}{2}} \Gamma (\Delta - d/2 + 3/2)},
\end{equation}
we compute the BCFT two-point function from the dictionary. However, since \eqref{b2bp} is identical to the boundary block $g^{b}_{\hat{\Delta}_{I}}$ from \eqref{blocks} up to a constant, one can rewrite the two-point function exclusively in terms of boundary data and the normalization constants $c_n$ of the modes $\psi_n$
\begin{equation}
    \langle \mathcal{O}(x) \mathcal{O}(y) \rangle = \frac{1}{C_{\hat{\Delta}, d}} \sum \limits_{n} c_{n}^{2} C_{\hat{\Delta}_{n},d-1} 2^{-2 \hat{\Delta}_{n}} g^{b}_{\hat{\Delta}_{n}}(\xi)\,,
\end{equation}
where
\begin{equation}
    1=  c_n^2 \int d^{d+1}x \sqrt{-g} \,\psi_n(\rho)^2 \Psi_n(x)^2\,.
\end{equation}

Comparing this expression with the RHS of \eqref{channels}, the BOE coefficients are given by
\begin{equation}
    \mathcal{B}_{n} = \frac{1}{2^{\hat{\Delta}_{n}}} \sqrt{\frac{C_{\hat{\Delta}_{n,d-1}}}{C_{\Delta,d}}}c_{n}\,.
\end{equation}

We are now in model-dependent territory, since unlike the spacing of boundary dimensions $\hat{\Delta}_n$, the asymptotics of the $c_n$'s depend on the details of the background. In Takayanagi's ETW model~\cite{Takayanagi:2011zk} and D1/D5 SUSY Janus~\cite{Chiodaroli:2016jod},  $c_n \sim n^{a/2}$ for some power $a$ that depends on details. However, the approximate singularity we are after is universal and does not depend on the precise power $a$. In the large $n$ limit the BOE coefficients become
\begin{equation}\label{largenB}
    B_{n} \approx 2^{-\frac{\pi n}{\phi_{b}}} n^{a}B\,,
\end{equation}
where we used \eqref{E:spacing}. The exponent $a$ is one that depends on both $\Delta$ and $d$, and the coefficient $B$ is independent of $n$. The main takeaway here is the appearance of $\phi_{b}$ from the regular spacing of $\hat{\Delta}_{n}$. Therefore, using \eqref{largenB} and \eqref{E:spacing}, the sum over boundary blocks is of the form
\begin{equation}
    \langle \mathcal{O}(x) \mathcal{O}(y) \rangle \propto \sum \limits_{n} n^{a} r^{\frac{n\pi}{\phi_{b}}},
\end{equation}
whose large $n$ behavior gives a polylogarithm of degree $a$. This polylogarithm has a singularity at $r = e^{2i\phi_b}$, near which
\begin{equation}
    \langle \mathcal{O}(x)\mathcal{O}(y)\rangle \propto \frac{1}{(r- e^{2i \phi_b})^{1-a}}\,,
\end{equation}
This is the singularity we were looking for, one that is sharp at leading order in the gravity approximation but (presumably) smoothed out by stringy corrections.\footnote{The authors of~\cite{Reeves:2021sab} have proven that such a singularity is at best approximate in 1+1 dimensions. The absence of such an exact singularity in a BCFT/DCFT with a stringy dual can be seen explicitly for certain conformal defects/boundaries in the context of the AdS$_3\times\mathbb{S}^3\times\mathbb{T}^4$ vacuum of IIB with a single unit of NS-NS flux~\cite{Gaberdiel:2021kkp,Harris:2025wak}.}

The argument above applies for $d+1$-dimensional duals to BCFTs. By the usual geometric focusing argument we expect it applies even with a nontrivially fibered internal space. For a DCFT the singularity appears instead at $r = e^{i\phi_b}$.

\section{Bottom-up models}\label{ToyModel}

In this Section we review the simplest class of bottom-up models for an AdS/BCFT correspondence: Einstein gravity with a negative cosmological constant in $d+1$ dimensions coupled to a tensionful ETW brane~\cite{Takayanagi:2011zk}. We also consider the related Karch-Randall model~\cite{Karch:2000ct} for AdS/DCFT~\cite{Karch:2000gx}. In this latter scenario we again have Einstein gravity coupled to a tensionful brane, which now serves as a domain wall between two asymptotically AdS$_{d+1}$ regions. In each of these models we recapitulate the light crossing time $\phi_b$ as a function of the tension $T$ of the ETW- or Karch-Randall brane, and review the boundary/defect entropy in the dual bottom-up BCFT/DCFT. The light crossing time was already found in~\cite{Reeves:2021sab} for Takayanagi's model and the entropy  in~\cite{Kobayashi:2018lil}. With the boundary/defect entropy as a function of $\phi_b$ in hand we will be equipped to compare with top-down models of DCFT and BCFT in Sections~\ref{Interfaces} and~\ref{Boundaries} respectively.

\subsection{Causal structure}

\subsubsection{ETW brane model}\label{S:ETW}

Takayanagi's model of BCFT~\cite{Takayanagi:2011zk} is described by the action
\begin{equation}\label{Tadashi}
    S = \frac{1}{16 \pi G} \int_{N} d^{d+1}x \sqrt{-g} \left(R + d(d-1)) \right) + \frac{1}{8 \pi G} \int_{Q}d^dx \sqrt{-h} (K - T) + (\text{terms on }M)\,,
\end{equation}
in  $-+..+$ signature. Here, the bulk spacetime is $N$ with metric $g$, and its boundary is the union of two segments: one, $M$ is the conformal boundary where the dual description ``lives,'' while the other $Q$ with induced metric $h$ is the worldvolume of the ETW brane. The parameter $T$ is the tension of the ETW brane, $K$ is the trace of the extrinsic curvature of the boundary, and $\Lambda = -\frac{d(d-1)}{2}$ is a negative cosmological constant in $L = 1$ units. In the putative dual BCFT we have in mind the vacuum in the presence of a flat boundary. The bulk geometry is then a cutout of AdS$_{d+1}$ described by~\eqref{MainMetric} with a warpfactor $A(\rho) = \ln(\cosh(\rho)).$ Explicitly, 
\begin{equation}\label{EmptyPoinAdS}
    ds^{2} = d\rho^{2} + \cosh^{2}(\rho) \left(\frac{-dt^{2} + d \mathbf{x}^{2} + dz^{2}}{z^{2}} \right)\,,
\end{equation}
for $\rho \geq \rho_0$ where the ETW brane is located at constant $\rho = \rho_0$. 

To find $\rho_0$ we solve the Israel junction conditions\,,
\begin{equation}
    K_{ab} - h_{ab} K = -T h_{ab}\,.
\end{equation}
Since $\rho$ is a Gaussian normal coordinate to the ETW brane we have
\begin{equation}
    K_{ab} = -\frac{1}{2} \partial_{\rho} h_{ab}\,,
\end{equation}
with the indices $a = t, \mathbf{x}, z$. From this, we find 
\begin{equation}\label{branetension}
    T = -(d-1) \, \tanh(\rho_{0})\,,
\end{equation}
which we invert to give
\begin{equation}\label{branelocation}
    \rho_{0} = -\text{arctanh} \left(\frac{T}{d-1} \right)\,.
\end{equation}
The tension is bounded as $|T| \leq d-1$ with positive tension corresponding to negative $\rho_0$ and vice versa.

We now compute $\phi_{b}$ using ~\eqref{PhiB}
\begin{equation}
    \phi_{b} = \int_{\rho_{0}}^{\infty} \frac{d\rho}{\cosh(\rho)} =\frac{\pi}{2} - 2 \, \text{arctan}\left(\tanh \left(\frac{\rho_{0}}{2} \right) \right) \,,
\end{equation}
and along the way we note a particularly useful relation between the brane's location $\rho_{0}$ and $\phi_{b}$
\begin{equation}
\label{E:ETWfromRhoToPhi}
	\tanh(\rho_0) = \cos(\phi_b)\,.
\end{equation}
Expressed as a function of $T$ we see that $\phi_b$ is a monotonically increasing function of $T$ valued in the domain $ [0,\pi]$ with the limits $\phi_b\to 0$ and $\phi_b\to \pi$, respectively; these correspond to the minimum and maximum possible values of the tension $T\to \pm (d-1)$. Moreover positive tension corresponds to $\phi_b>\pi/2$ and negative to $\phi_b<\pi/2$. Note that in the limit $\rho_0\to -\infty$, i.e. $\phi_b\to \pi$, the bulk geometry is nearly empty AdS$_{d+1}$. Further, a tensionless ETW brane, i.e. $\phi_b = \pi/2$, corresponds to the $\mathbb{Z}_2$ orbifold $\rho \sim -\rho$ of empty AdS.

\subsubsection{Karch-Randall braneworld}

Now consider the Karch-Randall (KR) model~\cite{Karch:2000ct} for the vacuum of a flat space DCFT with a flat defect~\cite{Karch:2000gx}. The dual geometry can be obtained by taking two AdS$_{d+1}$ cutouts like those above and gluing them together across their common boundary in such a way as to satisfy the Israel junction condition. The warpfactor is now $A(\rho) = \ln \cosh(|\rho|+\rho_0)$ with the KR brane located at $\rho=0$ and the junction condition reads
\begin{equation}
    \Delta\left(K_{ab}-K h_{ab}\right)=-Th_{ab} \,.
\end{equation}
The LHS describes the change in extrinsic curvature across the KR brane. Solving for the tension gives twice the result~\eqref{branetension} for an ETW brane of the same $\rho_0$,
\begin{equation}\label{KRbranetension}
    T= -2(d-1)\tanh\left(\rho_0\right) \,.
\end{equation}
Moreover $\phi_b$ is twice its value for a tensionful ETW brane of the same $\rho_0$, obeying
\begin{equation}
	\tanh(\rho_0) = \cos\left( \frac{\phi_b}{2}\right)\,.
\end{equation}
This immediately tells us that in the KR model $\phi_b$ is a monotonically increasing function of tension $T$ valued in the range $[0,2\pi]$. Negative tension KR branes correspond to $\phi_b\in [0,\pi]$, zero tension (i.e. empty AdS$_{d+1}$) to $\phi_b=\pi$, and positive tension KR branes to $\phi_b\in [\pi,2\pi]$. 

\subsection{Entropy}\label{ETWtoKR}

Having obtained the light crossing time $\phi_b$ we now obtain defect/boundary entropies associated with these bottom-up models focusing on the cases where the dual theory has dimension $d=2,3,4$. While there are many possible notions of defect or boundary entropy in $d > 2$ dimensions, we use one that reduces to the standard definition in $d=2$ of Affleck and Ludwig~\cite{Affleck:1991tk}. This definition decreases under defect/boundary RG flows in $d=2$~\cite{Friedan:2003yc} and $3$~\cite{Jensen:2015swa}, and there is some evidence that it does so in $d=4$~\cite{Estes:2014hka,Gaiotto:2014gha,Kobayashi:2018lil}. In this definition we consider the vacuum entanglement entropy $S_{\rm EE}$ across a sphere of radius $R$ centered on the defect or boundary, and subtract off a bulk contribution. In an equation,
\begin{align}
\begin{split}
\label{E:entropy1}
	\mathcal{S} = \begin{cases} S_{\rm EE} - S_{\rm CFT}\,, & \text{defect}\,, \\ S_{\rm EE} - \frac{1}{2}S_{\rm CFT}\,, & \text{boundary}\,,\end{cases}
\end{split}
\end{align}
where $S_{\rm CFT}$ is the vacuum EE of the CFT across a sphere of radius $R$. For a codimension-1 defect/boundary $\mathcal{S}$ behaves like the vacuum EE of a $d-1$-dimensional CFT. The defect/boundary entropy we study is simply the universal term in $\mathcal{S}$. For a defect in $d=2,3,4$ we have
\begin{align}
\begin{split}
\label{E:entropy2}
	\mathcal{S} = \begin{cases} D_0  + O(\varepsilon)\,, & d=2\,, \\ D_0 \ln \frac{R}{\varepsilon}  + \widetilde{D}_0 + O(\varepsilon)\,, & d=3\,, \\ \frac{a_1}{\varepsilon} - D_0  + O(\varepsilon)\,, & d=4\,, \end{cases}
\end{split}
\end{align}
where $\varepsilon$ is a short-distance regulator and $D_0$ refers to the defect entropy. For a boundary $\mathcal{S}$ has the same form only we label the universal term with $B_0$ instead of $D_0$. 

In $d=2$ the defect/boundary entropy is simply $ \ln g$, the entropy of Affleck and Ludwig~\cite{Affleck:1991tk}. In $d=3$ it is $\frac{b}{3}$ with $b$ the $a$-type defect/boundary central charge~\cite{Jensen:2015swa}. In $d=4$ it is the ``defect/boundary $F$''~\cite{Gaiotto:2014gha}.


In our conventions the authors of~\cite{Kobayashi:2018lil} have previously computed the boundary entropy $B_0$ in Takayanagi's model to be
\begin{equation}\label{universalbottomupentropy}
    B_{0}= \frac{1}{4G}\frac{\pi^{(d-1)/2}}{\Gamma\left(\frac{d-1}{2}\right)} \tanh( \rho_0)\,_2F_1\left(\frac{1}{2},\frac{d}{2},\frac{3}{2},\tanh^2\rho_0\right)\times \begin{cases} 1 \,, & d\text{ even}\,, \\ \frac{2}{\pi}\,, & d \text{ odd}\,. \end{cases}
\end{equation}
In terms of $\phi_b$ we can use~\eqref{E:ETWfromRhoToPhi} to replace $\tanh(\rho_0)$ with $\cos(\phi_b)$. This entropy is monotonically increasing from $B_0\to -\infty$ as $\phi_b\to 0$ and $B_0\to \infty$ as $\phi_b\to \pi$, with negative $B_0$ corresponding to $\phi_b$ in the range $[0,\pi/2]$ and positive to the range $[\pi/2,\pi]$. In terms of tension this means that $B_0$ is negative for a negative tension ETW brane and positive for a positive tension ETW brane. 

A similar computation in~\cite{Kobayashi:2018lil} gives $D_0 = 2B_0$ at the same value of $\rho_0$. By the discussion of the previous Subsection we can replace $\tanh(\rho_0) = \cos\left( \frac{\phi_b}{2}\right)$. Viewed as a function of $\phi_b$ the defect entropy also monotonically increases from $D_0\to -\infty$ as $\phi_b\to 0$ to $D_0\to \infty$ as $\phi_b\to 2\pi$. It is negative now in the range $[0,\pi]$ and positive in $[\pi,2\pi]$. Again, $D_0$ has the same sign as that of the brane tension.

\section{Top-down defects}
\label{Interfaces}

In this Section we compute $\phi_{b}$ for a variety of top-down holographic DCFTs and describe the defect entropy $D_0$ as a function of $\phi_b$. We study the following top-down defects: non-SUSY Janus in 3+1 dimensions; a SUSY Janus deformation of the D1/D5 system; the  D3/D5 intersection; a SUSY Janus of $\mathcal{N}=4$ super Yang-Mills (SYM); and M-theory Janus. We then compare with the bottom-up Karch-Randall model reviewed in the previous Section. In all of our examples $\phi_b$ obeys the bound $\phi_b\geq \pi$ and the defect entropy is a positive, monotonically increasing function of $\phi_b$.

Before we begin, we note that in the Appendix we also consider a class of pseudo-bottom-up models in the form of a massless scalar field added to a general ($d+1$)-dimensional Einstein gravity with negative cosmological constant. This latter example provides an interesting ``bridge'' to our top-down constructions. In particular, for a spacetime dimension of $d + 1 = 4+1$ this solution is precisely that dual to the non-SUSY Janus deformation of $\mathcal{N}=4$ SYM.

\subsection{Causal structure}
\subsubsection{Non-SUSY Janus}\label{nonSUSYJanusSec}

The non-SUSY Janus solution of type IIB SUGRA is a one-parameter deformation of the AdS$_{5} \times \mathbb{S}^{5}$ solution in which only the 5d metric, dilaton, and RR 5-form are non-trivial. The resulting solution breaks all supersymmetry of the AdS$_5\times\mathbb{S}^5$ vacuum. The Einstein-frame metric of this solution is
\begin{equation}
\label{E:metricnonSUSYJanus}
    ds^{2} = L^{2} \left( \gamma^{-1} h(\xi)^{2} d \xi^{2} + h (\xi) ds^2_{\text{AdS}_{4}} \right) + L^{2} ds_{\mathbb{S}^{5}}^{2},
\end{equation}
in which $L^{4} = 4 \pi N_{3} \alpha'^{2}$. Here $\gamma$ is a real parameter which parameterizes the difference of complexified Yang-Mills couplings on either side of the Janus interface and obeys $\frac{3}{4} \leq \gamma \leq 1$~\cite{Bak:2003jk,DHoker:2006vfr}. The lower bound gives a linear dilaton solution, whereas $\gamma = 1$ describes empty AdS$_5\times\mathbb{S}^5$. The  warpfactor is given by
\begin{equation}
    h(\xi) = \gamma \left(1 + \frac{4 \gamma - 3}{\mathfrak{p}(\xi) + 1 - 2 \gamma} \right)\,,
\end{equation}
where $\mathfrak{p}(\xi)$ is the Weierstrauss elliptic function, defined by the equation
\begin{equation}
    (\partial_{\xi} \mathfrak{p})^{2} = 4 \mathfrak{p}^{3} - g_{2} \mathfrak{p} - g_{3}\,,
\end{equation}
with periods $g_{2} = 16 \gamma (1 - \gamma)$ and $g_{3} = 4 (\gamma - 1)$.

\begin{figure}[t]
    \centering
    \includegraphics[width=3.5in]{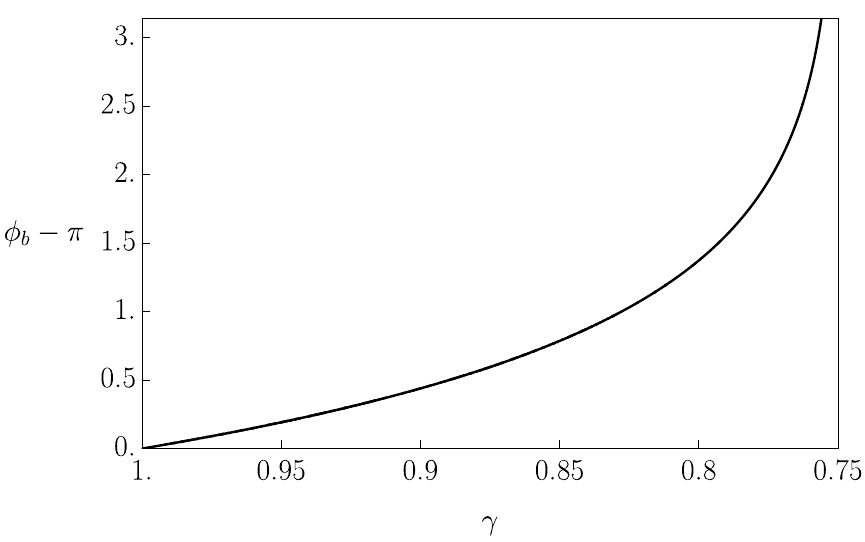}
    \caption{\label{nonSUSYFig}The light crossing time $\phi_{b}$ as a function of the parameter $\gamma$. The limit $\gamma=1$ corresponds to empty AdS.   }
\end{figure}

We now compute $\phi_{b}$. The shortest trajectories in this geometry sit at a constant angle on the $\mathbb{S}^5$ with motion entirely in the 5d part of the geometry,
\begin{equation}
    ds^{2}_5= \gamma^{-1} h(\xi)^{2} d \xi^{2} + h(\xi) \left(\frac{-dt^{2} + d z^{2} + d\mathbf{x}^{2}}{z^{2}} \right)\,,
\end{equation} 
of the form~\eqref{MainMetric} under the change of coordinates $d\rho = \gamma^{-1/2}h(\xi)d\xi$. Thus
\begin{equation}
    \phi_{b} = \int^{\infty}_{-\infty} d\xi\sqrt{\frac{h(\xi) }{\gamma}}\,,
\end{equation}
which we evaluate numerically as a function of $\gamma$. See Fig.~\ref{nonSUSYFig}. For all $\gamma$ we have $\phi_b\geq \pi$; at $\gamma = 1$, we recover the empty AdS result $\phi_{b} = \pi$, with decreasing $\gamma$ corresponding to increasing $\phi_b$. The limit $\gamma \to \frac{3}{4}$ is finite. This is the only top-down example in which $\phi_b$ is upper-bounded.  

\subsubsection{D1/D5 SUSY Janus}

Next we consider SUSY Janus deformations of the AdS$_3\times\mathbb{S}^3\times M_4$ vacuum of IIB SUGRA with $M_4=\mathbb{T}^4$ or $K3$, i.e. of the D1/D5 system. These deformations do not alter the $M_4$ and so can be described in terms of a six-dimensional Einstein frame metric~\cite{Chiodaroli:2010ur}
\begin{equation}\label{reducedD1D5}
    ds^2_6=f_1^2 ds^2_{\text{AdS}_2}+f_2^2 ds^2_{S^2}+f_{3}^2 (dx^2+dy^2) \,,
\end{equation}
in which the $x$-coordinate will behave as an effective radial coordinate. The warpfactors are 
\begin{align}
\begin{split}
    H^2&=f_1^2 f_2^2 = 4 L^2 \cosh^2(x)\sin^2(y)\,,
     \\
    \frac{f_{3}^2}{f_1^2}&=\frac{\kappa^2}{\cosh^2{x}}\,,
    \\
    \frac{f_{3}^2}{f_2^2}&=\frac{1}{\sin^2{y}}+\frac{\kappa^2-1}{\cosh^2{x}}\,,
\end{split}
\end{align}
where $\kappa=\cosh{\psi}\cosh{\theta}\geq 1$ is a parameter labeling the solution. More explicitly,
\begin{align}
\begin{split}
    f_1^4&=\frac{4 L^2 }{\kappa ^2} \cosh ^2(x) \left(\cosh ^2(x)+\left(\kappa ^2-1\right) \sin ^2(y)\right)\,,
    \\
    f_2^4&=\frac{4 \kappa ^2 L^2 \cosh ^2(x) \sin ^4(y)}{\cosh ^2(x)+\left(\kappa ^2-1\right) \sin ^2(y)}\,,
     \\
    f_{3}^4&=\frac{4 \kappa ^2 L^2}{\cosh ^2(x)} \left(\cosh ^2(x)+\left(\kappa ^2-1\right) \sin ^2(y)\right)\,.
\end{split}
\end{align}
The two halves of the $1+1$-dimensional conformal boundary are reached as $x\to \pm \infty$. A priori it is then unclear which trajectory corresponds to the light crossing time. However, we are fortunate that $\phi_b$ can also be computed from the two-point function of DCFT operators. As emphasized by~\cite{Reeves:2021sab} the authors of~\cite{Chiodaroli:2016jod} have computed the two-point function of a certain scalar half-BPS operator, from which they extracted $\phi_b$ with the result
\begin{equation}
	\phi_b = \pi \kappa\geq \pi\,,
\end{equation}
which we note always exceeds $\pi$ just as $\phi_b$ in the non-SUSY Janus solution studied above.  In this Subsection we reproduce this result simply from the 6d geometry.  

In particular we claim that the shortest time-of-flight trajectories from one half of the conformal boundary to the other reside at $y=0,\pi$, for which the motion is effectively three-dimensional, taking place in the AdS$_2$ fibers and the $x$-direction, and that the light-crossing time for these trajectories is the value $\pi\kappa$ obtained from DCFT correlations.

Let us first verify that setting $y=0,\pi$ is consistent with the geodesic equation. Near $y=0$ 
\begin{align}
\begin{split}
    f_1^2&=f_1^{(0)}(x)+f_1^{(2)}(x)y^2+O(y^3) \,,\\
    f_2^2&=f_2^{(2)}y^2+O(y^3) \,,\\
    f_{3}^2&=f_{3}^{(0)}+f_{3}^{(2)}(x)y^2+O(y^3) \,,
\end{split}
\end{align}
where $f_2^{(2)}$ and $f_{3}^{(0)}$ are constants. Because the deviations are quadratic in $y$ there is no effective force on a particle at $y=0$, so that $y=0$ is indeed consistent. The 6d geometry has a symmetry under $y\to \pi - y$ and so the same statement holds for $y=\pi$. Now, at $y=0,\pi$, the $\mathbb{S}^2$ shrinks to zero which simplifies the effective metric in which the motion takes place to be
\begin{equation}
   ds^2_{\rm eff} = 2L\kappa\,dx^2 + \frac{2L}{\kappa}\cosh^2(x)ds^2_{\text{AdS}_2}\,.
\end{equation}
From this we see that the physical radius $L_{\rm eff}^2$ of the asymptotically AdS$_3$ regions is $L_{\rm eff}^2 = 2L\kappa^2$. We then compute 
\begin{equation}
    \phi_b=\int_{-\infty}^\infty dx \frac{\kappa}{\cosh{x}} = \kappa \pi \,,
\end{equation}
reproducing the result above.

\subsubsection{D3/D5 intersection}\label{D3D5Subsec}

Another common DCFT is 3+1-dimensional $\mathcal{N} = 4$ SYM theory coupled to $N_{5}$ $2+1$-dimensional hypermultiplets living on a flat defect in such a way as to preserve eight supercharges of SUSY. This type of DCFT appears as the worldvolume theory living on the D3/D5 intersection. Letting $z$ refer to the direction perpendicular to the defect, we can take the gauge group to be $SU(N_3^+)$ for $z>0$ and $SU(N_3^-)$ for $z<0$, in which case the corresponding brane intersection has a stack of coincident D3 branes intersecting across a stack of $N_5$ coincident D5 branes, with $N_3^{\pm}$ labeling the number of D3 branes on either side of the D5s. 

In the large $N_3^{\pm}$ and $N_5$ limits this brane intersection supports a near-horizon geometry of IIB SUGRA which is locally AdS$_4\times\mathbb{S}^2\times\mathbb{S}^2\times \Sigma$ with $\Sigma$ a strip. Parameterizing the Einstein-frame metric by
\begin{equation}\label{N4metric}
    ds_{10}^2=f_4^2 ds_{\text{AdS}_4}^2+f_{3}^2 (dx^2+dy^2)+f_1^2 ds_{\mathbb{S}^2}^2+f_2^2 ds_{\bar{\mathbb{S}}^2}^2 \,,
\end{equation}
the warpfactors are given by~\cite{Gomis:2006cu,DHoker:2007zhm}
\begin{align}\label{N4VolumeFactors}
\begin{split}
    f_4^8&=16\frac{F_1 F_2}{w^2} \,, \hspace{.96in}f_{3}^8=\frac{2^8 F_1 F_2 w^2}{h_1^4 h_2^4}\,, \\
    f_1^8&=16 h_1^8\frac{F_2 w^2}{F_1^3} \,, \hspace{.8in}  f_2^8=16 h_2^8\frac{F_1 w^2}{F_2^3} \,, \\
    F_i&=2 h_1 h_2|\partial_v h_i|^2-h_i^2 w \,,\hspace{.27in} w=\partial_v\partial_{\bar{v}}(h_1 h_2) \,,
\end{split}
\end{align}
where $v=x+iy$ is the complex coordinate on the strip with $x\in (-\infty,\infty)$ and $y\in[0,\pi/2]$, and $h_1,h_2$ are harmonic functions on the strip. For the D3/D5 intersection they are given by~\cite{DHoker:2007hhe}
\begin{align}
\begin{split}
\label{E:hD3D5}
    h_1(v,\bar{v})&=\alpha'\left(-i\alpha \sinh\left(v\right)-\frac{N_5}{4}\ln{\left(\tanh\left(\frac{i\pi}{4}-\frac{v-\delta}{2}\right)\right)}\right)+(\text{c.c.}) \,, \\
    h_2(v,\bar{v})&=\alpha' \hat{\alpha} \cosh\left(v\right)+(\text{c.c.}) \,,
\end{split}
\end{align}
where $\alpha, \delta, \hat{\alpha}$ are fixed by $g_{\rm YM}, N_3^{\pm}, N_5$ via
\begin{align}
\begin{split}
\label{D3D5parameters}
    \alpha &= -\frac{N_{5}}{4} \cosh(\delta) + \sqrt{\frac{\pi^{2} N_{3}}{g^2_{\rm YM}} + \frac{N_{5}^{2}}{16} \cosh^{2}(\delta)}\,, \qquad \hat{\alpha} = g^2_{\rm YM} \frac{\alpha}{4 \pi}\,,
	\\
    e^{\delta} &= \sqrt{\frac{2 g^2_{\rm YM} N_{3} N_{5}^{2} + 4 \pi^{2} \Delta N_{3}^{2} + \sqrt{(2 g^2_{\rm YM} N_{3} N_{5}^{2} + 4 \pi^{2} \Delta N_{3}^{2})^{2} - g^{4}_{YM} N^{4}_{5} (4 N_{3}^{2} - \Delta N_{3}^{2})}}{g^2_{\rm YM} N^{2}_{5} (2 N_{3} - \Delta N_{3})}}\,,
\end{split}
\end{align}
where $N_{3} \equiv \frac{1}{2} (N_{3}^{+} + N_{3}^{-})$ and $\Delta N_3 = N_3^+-N_3^-$. The limit $\Delta N_3=0$ corresponds to $\delta = 0$. We take $\delta$ and  $N_{5} / \alpha$ to be $\mathcal{O}(1)$ with $g_{\rm YM}\to 0$ in which case the 10d geometry is weakly curved and gravity is semiclassical.

\begin{figure}[t]
    \centering
    \includegraphics[width=3.5in]{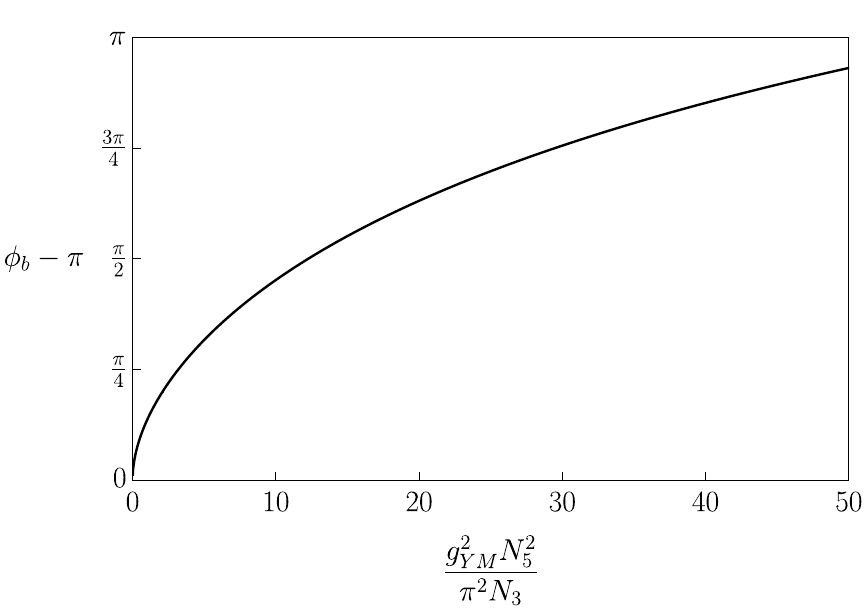}
    \caption{\label{D3D5Fig} The light crossing time $\phi_{b}$ for the D3/D5 DCFT when the number of D3 branes does not jump across the defect.}
\end{figure}

The two halves of the conformal boundary are located as $x\to \pm \infty$. In studying null trajectories connecting them we can restrict the motion to take place only in the AdS$_4$ and strip directions. As in our discussion of the D1/D5 Janus solution, we can look for motion that is effectively one-dimensional on the strip, fixed to a constant value of $y$. A simple exercise shows that fixing $y=0,\pi/2$ are choices that are consistent with the geodesic equation, with $y=0$ corresponding to a finite light crossing time and $\pi/2$ to an infinite one. The dominant one contributing to singularities in boundary correlation functions will then be the $y=0$ trajectory, for which the light-crossing integrand is
\begin{equation}\label{D3D5IntIntegrand}
    e^{-A(\rho)} \, d \rho=  \frac{f_3}{f_4} \, dx =  \text{sech}(x) \sqrt{\frac{N_5 \cosh (2 x-\delta ) \text{sech}(x-\delta )+4 \alpha  \cosh (x-\delta )}{N_5 \text{sech}(x)+4 \alpha  \cosh (x-\delta )}} \, dx \,,
\end{equation}
which we can numerically integrate as a function of fixed $\frac{N_5}{\alpha}$ and $\delta$. We find that $\phi_b$ always obeys
\begin{equation}
	\phi_b\geq \pi\,.
\end{equation}
In order to compare with the Karch-Randall model, where the 5d cosmological constant does not jump across the KR brane, we mostly keep in mind the case $\delta=0$, where the number of D3 branes (and thus the radius of the asymptotically AdS$_5$ regions) does not jump. In this case we plot $\phi_b$ as a function of field theory parameters in Fig.~\ref{D3D5Fig} using~\eqref{D3D5parameters}.

In the next Section we will take the limit $N_3^- \to 0$ in which case this DCFT becomes the BCFT describing D3 branes ending on D5 branes. That case corresponds to a certain scaling limit in which $\delta\to\infty$. Naively we can extract $\phi_b$ for the BCFT from the large $\delta$ limit of $\phi_b$ for the DCFT, however we will see that this is not the case and only a part of the light-crossing integrand above contributes to the light crossing time of the BCFT. This leads to a regime of BCFT parameter space where $\phi_b<\pi$. 

\subsubsection{SYM SUSY Janus}

\begin{figure}[t]
    \centering
    \includegraphics[width=3.5in]{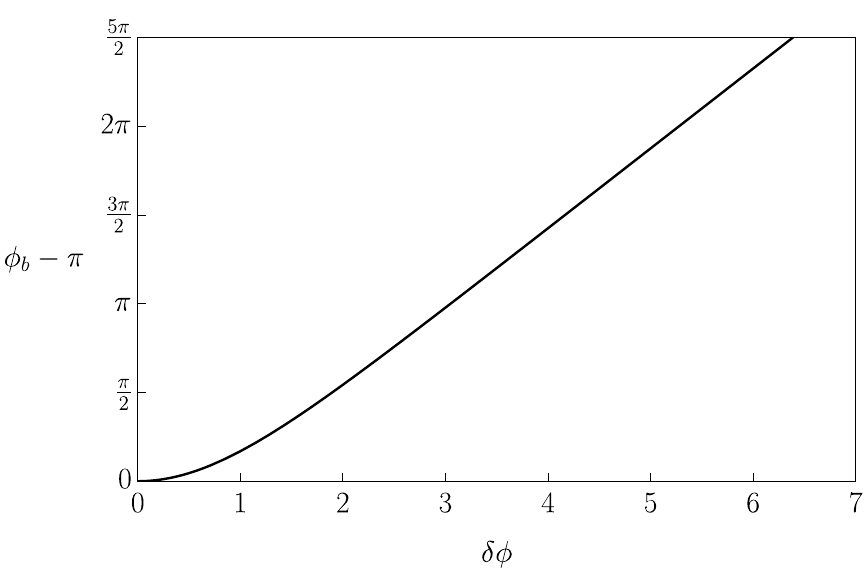}
    \caption{\label{SUSYJanusFig} The light crossing time of the SUSY Janus deformation of $\mathcal{N}=4$ SYM as a function of $\delta \phi$. The light crossing time tends to infinity at large $\delta\phi$.}
\end{figure}

We next consider the SUSY Janus deformation of $\mathcal{N}=4$ SYM. Like the dual of the D3/D5 intersection, the dual geometry is also locally AdS$_4\times\mathbb{S}^2\times\mathbb{S}^2\times \Sigma$ described by~\eqref{N4metric} but now the harmonic functions are given by~\cite{DHoker:2007zhm}
\begin{equation}
    h_1(v,\bar{v})=-i\alpha_1 \sinh\left(v-\frac{\delta\phi}{2}\right)+(\text{c.c.}) \,, \qquad  h_2(v,\bar{v})=\alpha_2 \cosh\left(v+\frac{\delta\phi}{2}\right)+(\text{c.c.})\,,
\end{equation}
where $(\alpha_1,\alpha_2,\delta\phi)$ are real constants related to the asymptotic radius of curvature $L$ and the Yang-Mills couplings (in a $SL(2;\mathbb{R})$ duality frame in which the theta-angle is constant) $g^{\pm}_{\rm YM}$ on either side of the interface are
\begin{equation}
    L^{4} = 16 |\alpha_{1} \alpha_{2}| \cosh(\delta \phi)\,, \qquad \frac{(g^{\pm}_{\rm YM})^{2}}{4 \pi} = \bigg\vert\frac{\alpha_{2}}{\alpha_{1}} \bigg\vert e^{\pm \delta \phi}\,,
\end{equation}
The parameter $\delta\phi$ parameterizes the strength of the Janus deformation. We rewrite the harmonic functions more simply as
\begin{equation}
    h_1(x,y)=2\alpha_1 \cosh\left(x-\frac{\delta\phi}{2}\right)\sin(y) \,, \quad  h_2(x,y)=2\alpha_2 \cosh\left(x + \frac{\delta\phi}{2}\right)\cos(y) \,,
\end{equation}
which we use to determine the warpfactors given in \eqref{N4VolumeFactors}. Similar to the D3/D5 intersection, we consider geodesic motion at fixed angles on the two $\mathbb{S}^2$'s and look for simple motion on the strip. We note that $f_{1}$ vanishes at $y=0$ while $f_{2}$ vanishes at $y=\pi/2$, with both these locations again consistent with the geodesic equation. The light crossing time for both trajectories is identical with the integrand being
\begin{equation}
     e^{-A(\rho)} \, d \rho =\frac{f_{3}}{f_{4}} \, dx = \frac{\sqrt{2}}{\sqrt{\text{sech}(\delta \phi ) \cosh (2 x)+1}} \, dx \,,
\end{equation}
and therefore the light crossing time is
\begin{equation}
    \phi_b = \frac{2\sqrt{\cosh(\delta\phi)}}{\cosh\left( \frac{\delta\phi}{2}\right)}K\left(\tanh^2\left( \frac{\delta\phi}{2}\right)\right) \geq \pi\,,
\end{equation}
where $K$ is the complete elliptic integral of the first kind. We plot this result in Fig.~\ref{SUSYJanusFig}.

\subsubsection{M-theory Janus}\label{MJanus}

Our last example is the M-theory Janus solution~\cite{DHoker:2009lky,Bobev:2013yra,Kim:2009wv, Bachas:2013vza,Estes:2014hka}. The original version of the solution is a one parameter deformation of the AdS$_{4} \times \mathbb{S}^{7}$ solution of eleven-dimensional SUGRA which preserves half the SUSY. For now we consider a generalization~\cite{Bachas:2013vza} in which there is a second parameter $\gamma >0$ (not to be confused with the $\gamma$ of our non-SUSY Janus solution of Subsection~\ref{nonSUSYJanusSec}) for which the supergravity background preserves a $D(2,1;\gamma,0)\times D(2,1;\gamma,0)$ supergroup. When the parameter $\gamma=1$ these solutions become the original M-theory Janus background and there is a dual interpretation in terms of ABJM theory at level $k=1$ deformed by certain localized operators along a flat defect. This solution can also be orbifolded~\cite{Estes:2014hka} to produce a dual to a Janus deformation of ABJM at generic level $k$.

The eleven-dimensional metric for this two-parameter deformation is locally AdS$_3\times\mathbb{S}^3\times\mathbb{S}^3\times \Sigma$ with $\Sigma$ a strip,
\begin{equation}\label{M2M5Intersection}
    ds^2=f_{4}^2 ds^2_{\text{AdS}_3} + f_{3}^2 \left(\frac{dx^2}{4}+dy^2\right) + f_1^2 ds^2_{\mathbb{S}^3}+f_2^2 ds^2_{\bar{\mathbb{S}}^3} \,,
\end{equation}
with $w=x/2+iy$ a complex coordinate on the strip with $x\in (-\infty,\infty)$ and $y\in[0,\pi/2]$. The warpfactors are 
\begin{align}
\begin{split}
    f_{4}^{6} &= \frac{h^2 W_+ W_-}{C_1^6(H\bar{H}-1)^2} \,, \hspace{.53in} f_{3}^{6} = \frac{|\partial_w h|^6}{C_2^3 C_3^3 h^4}(H\bar{H}-1) W_+ W_- \,, \\
    f_1^{6} &= \frac{h^2(H\bar{H}-1)W_-}{C_2^3 C_3^3 W_+^2} \,, \hspace{.4in} f_{2}^6=\frac{h^2(H\bar{H}-1)W_+}{C_2^3 C_3^3 W_-^2} \,,
\end{split}
\end{align}
with $C_1,C_2,C_3$ real constants satisfying $\gamma = \frac{C_{2}}{C_{3}}$and $C_1+C_2+C_3=0$, and
\begin{align}
\begin{split}
    h&=-2i\alpha\left(\sinh(2w)-\sinh(2\bar{w})\right) = 4\alpha \cosh(x) \sin(2y) \,,\\
    H &= i\frac{\cosh(w+\bar{w})+\lambda\sinh(w-\bar{w})}{\cosh(2\bar{w})} = \frac{i\cosh(x)-\lambda\sin(2y)}{\cosh(x)\cos(2y)-i\sinh(x)\sin(2y)} \,, \\
     W_\pm&=|H\pm i|^{2} + \gamma^{\pm 1}(H\bar{H}-1) \,.
\end{split}
\end{align}
The parameter $\lambda$ specifies the strength of the deformation, with $\gamma=1$ and $\lambda=0$ corresponding to the AdS$_4\times\mathbb{S}^7$ vacuum of M-theory. For $\gamma\neq 1$ the warpfactors $f_1$ and $f_2$ vanish as $y\to 0,\pi/2$, while $f_3$ and $f_4$ diverge. We henceforth consider $\gamma=1$ where the warpfactors are smooth at the boundaries of the strip and there is a known dual interpretation.

In computing $\phi_b$ we can again restrict to fixed angles on the $\mathbb{S}^3$'s and look for motion at fixed $y$. In this case both $y=0,\pi/2$ correspond to geodesic trajectories with identical light crossing time. The light crossing integrand is
\begin{equation}
    e^{-A (\rho)} \, d \rho= \frac{dx}{2}\frac{f_{3}}{f_{4}} = \frac{\sqrt{1+\lambda^2}}{\cosh{x}} \, dx \,.
\end{equation}
and thus
\begin{equation}
    \phi_b = \sqrt{1+\lambda^2}\,\pi \geq\pi \,.
\end{equation}
Again, the limit $\lambda \to 0$ corresponds to that of empty AdS$_4\times\mathbb{S}^7$.

\subsection{Defect entropy and comparing with Karch-Randall}\label{DefectComps}

The defect entropies $D_0$ are known for all of the examples we presented above. We recapitulate these results in the Appendix. Having computed $\phi_b$ in the last Subsection, we may then express the defect entropies of these examples as functions of $\phi_b$ and compare with $D_0$ as a function of $\phi_b$ in the Karch-Randall model. 

\begin{figure}[t]
	\centering
	\begin{minipage}{0.45\textwidth}
 		\centering
		\includegraphics[width=3in]{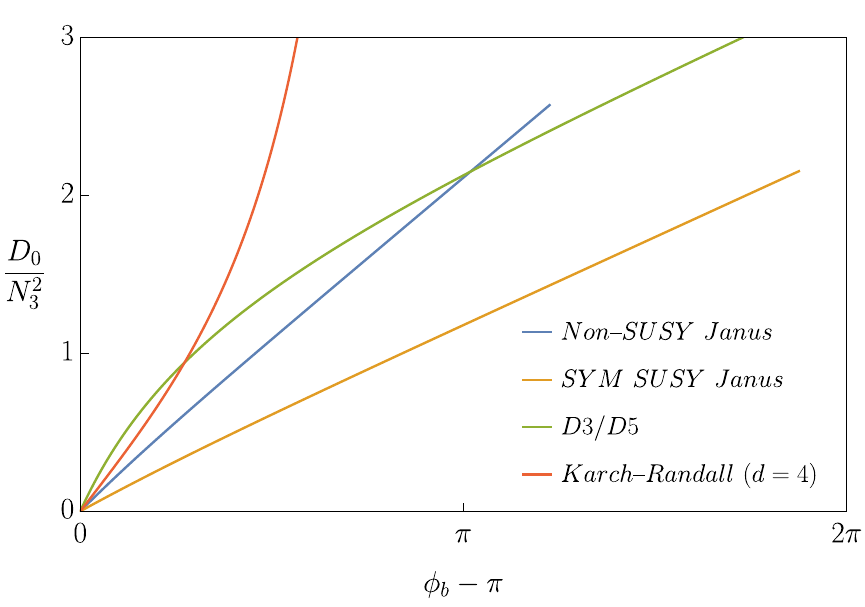}
		\caption{\label{d4DCFTCompFig} Plots of the normalized defect entropy $D_0/N_3^2$ in $d=4$ as a function of $\phi_b$. Note that the top-down defects only exist for $\phi_b\geq \pi$; here we plot the result for KR branes with $\phi_b\geq \pi$, equivalently positive tension. }
	\end{minipage}
	\hfill
	\begin{minipage}{0.45\textwidth}
   		\centering
		\includegraphics[width=3in]{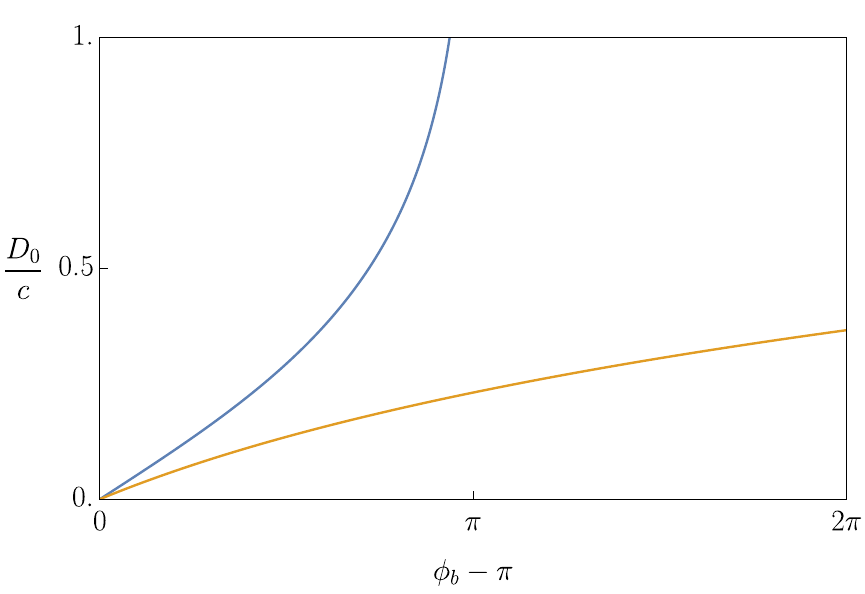}
		\caption{\label{d2DCFTCompFig}  A comparison of normalized defect entropy in $d=2$ between D1/D5 SUSY Janus and the Karch-Randall model. \\ \phantom{.}\\ \phantom{.}}
	\end{minipage}
\end{figure}

Only one thing remains. In comparing the bottom-up and top-down models we must ensure that we do so at identical values of the asymptotic AdS radius, or equivalently, of the lower-dimensional Newton's constant describing the effective theory on AdS far from the domain wall. In $d=2$ we can do this by normalizing $D_0$ by the central charge $c = \frac{3L}{2G}$ of the dual; in $d=3$ it turns out that the defect entropy of M-theory Janus vanishes~\cite{Estes:2014hka}, obviating the need for a comparison; and in $d=4$ we can trade the 5d Newton's constant $G_5$ for the number $N_3$ of D3 branes using
\begin{equation}
\label{E:d4MatchingG}
	\frac{L^8\text{vol}(\mathbb{S}^5)}{16\pi G_{10}} = \frac{L^8\pi^3}{(2\pi)^7\ell_p^8} = \frac{L^3}{16\pi G_5}\,, \qquad \Rightarrow \qquad \frac{L^3}{G_5} = \frac{2}{\pi}N_3^2\,,
\end{equation}
and then compare the normalized quantity $D_0/N_3^2$. Our $d=4$ results are summarized in Fig.~\ref{d4DCFTCompFig} and our $d=2$ results in Fig.~\ref{d2DCFTCompFig}. 

In each case the defect entropy is positive, monotonically increasing with $\phi_b$, with the entropy vanishing at $\phi_b=\pi$ where in each case the defect disappears. The qualitative behavior of the top-down and KR models match at the $\mathcal{O}(1)$ level, at least for positive tension KR branes. All defects have $\phi_b\geq \pi$ and therefore we find no top-down analogue of a negative tension KR brane. There are however top-down defects with $\phi_b>2\pi$ which cannot be imitated with the KR model.

\section{Top-down BCFTs}\label{Boundaries}

We turn our attention to top-down examples of holographic BCFTs with the same goals as in our discussion of top-down DCFTs in Section~\ref{Interfaces}. In this Section we study the BCFTs describing the worldvolume theory of D3 branes ending on D5 branes, and M2 branes ending on M5 branes, and then compare with Takayanagi's bottom up model.

\subsection{Causal structure}
\subsubsection{D3/D5 BCFT}

In Subsection~\ref{D3D5Subsec} we discussed the 10d geometry dual to the D3/D5 DCFT. The BCFT comes from taking the limit $N_3^-\to 0$. The 10d Einstein-frame metric is given by~\eqref{N4metric},~\eqref{N4VolumeFactors} where the harmonic functions $h_1$ and $h_2$ are given by
\begin{align}
\begin{split}
\label{E:hD3D5b}
	h_1(v,\bar{v}) & =-\alpha'\left(\frac{i\underline{\alpha}}{2}e^v + \frac{N_5}{4}\ln\tanh\left( \frac{i\pi}{4}- \frac{v}{2}\right)\right) + (\text{c.c.}) \,,
	\\
	h_2(v,\bar{v}) & = \alpha'\frac{\hat{\underline{\alpha}}}{2} e^v + (\text{c.c.})\,,
\end{split}
\end{align}
where $(\underline{\alpha},\hat{\underline{\alpha}})$ are constants labeling the solution. One way to arrive at this solution is to take a scaling limit of the D3/D5 DCFT in which the parameters $(\alpha,\hat{\alpha},\delta)$ are scaled as $\delta\to\infty$, $\alpha\to 0$ and $\hat{\alpha}\to 0$ while holding fixed
\begin{equation}
\label{E:hattedAlphas}
	\alpha e^{\delta} = \underline{\alpha} = \frac{2\pi^2 N_3}{g_{\rm YM}^2 N_5}\,, \qquad \frac{\hat{\alpha}}{\alpha} = \frac{\hat{\underline{\alpha}}}{\underline{\alpha}} = \frac{g_{\rm YM}^2}{4\pi}\,,
\end{equation}
where the RHS of the first equation is the limit of the LHS. Under this limit the harmonic functions $h_1$ and $h_2$~\eqref{E:hD3D5} describing the DCFT become those~\eqref{E:hD3D5b} provided that we first shift $x \to x+\delta$ and then take the limit. The ensuing geometry is weakly curved and gravity is semiclassical for $\underline{\alpha}$ of $\mathcal{O}(1)$ and $g_{\rm YM}^2 \to 0$.

\begin{figure}[t]
    \centering
    \includegraphics[width=3.5in]{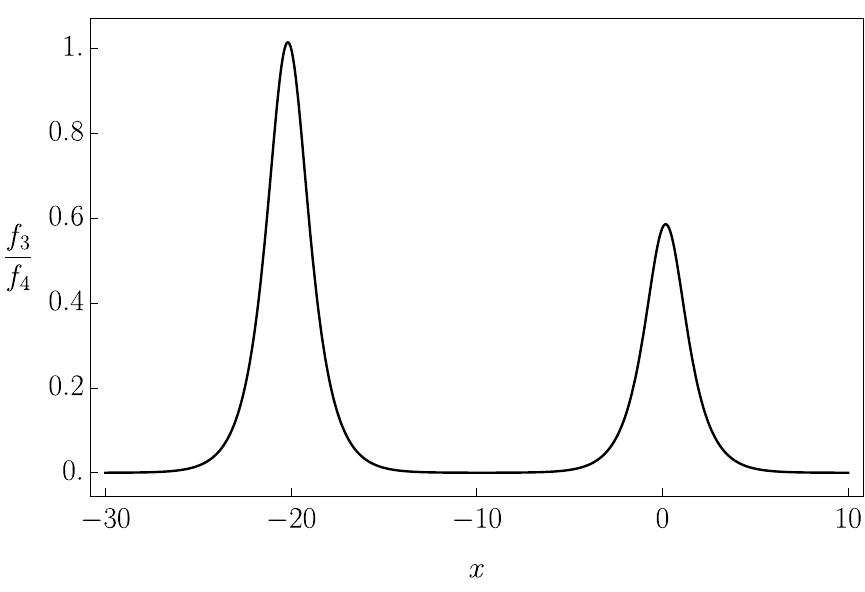}
    \caption{\label{BCFTLimitFig} The light crossing integrand of the D3/D5 DCFT for $\delta = 20$ and $\alpha e^{\delta}/N_5= 1$. In the BCFT limit $\delta\to \infty$ the large peak at $x=-\delta$ corresponds to a spurious spacetime region of vanishing size $ds^2\propto e^{-\delta}$ and does not contribute to the BCFT light crossing time.}
\end{figure}

Naively we can obtain the light crossing time by simply taking the scaling limit of the DCFT light crossing time. However this is not the case. To understand why see Fig.~\ref{BCFTLimitFig}. There we have plotted the light crossing integrand at reasonably large $\delta$ and having shifted $x\to x+\delta$. The integrand has two peaks, one centered around $x=-\delta$, and another centered around $x=0$. The former corresponds to a region of the 10d spacetime that disappears in the BCFT limit and does not contribute to any physical quantity. It is a region of vanishing size in which $ds_{10}^2 \propto e^{-\delta}$ but in which the ratios of warpfactors, like that $\frac{f_3}{f_4}$ that determine the light-crossing integrand, remain finite. The BCFT light crossing time is then the scaling limit of the DCFT light crossing time after subtracting off this spurious contribution. 

Alternatively we can construct the light crossing integrand directly from the geometry described above. The geodesic sits at a constant angle on the $\mathbb{S}^2$'s and at $y=0$, so that
\begin{equation}
    d \rho \, e^{-A(\rho)}= dx \, \frac{f_{3}}{f_{4}}= dx \, e^{x} \sqrt{\frac{N_5(1- \tanh (x))}{\underline{\alpha} (1+ e^{2 x})+N_5}} \,,
\end{equation}
which we integrate to obtain
\begin{equation}\label{phibD3D5BCFT}
    \phi_{b} =\sqrt{\frac{2N_5}{\underline{\alpha}}}K\left(-\frac{N_5}{\underline{\alpha}}\right) \,,
\end{equation}
where $K$ is the complete elliptic integral of the first kind. In terms of field theory parameters, this expression is
\begin{equation}
    \phi_{b} = \sqrt{\frac{g^{2}_{YM}N^{3}_{5}}{ \pi^{2} N_{3}}}K\left(-\frac{g^{2}_{YM} N^{3}_{5}}{2 \pi^{2} N_{3} }\right)\,.
\end{equation}
    
\begin{figure}[t]
    \centering
    \includegraphics[width=3.5in]{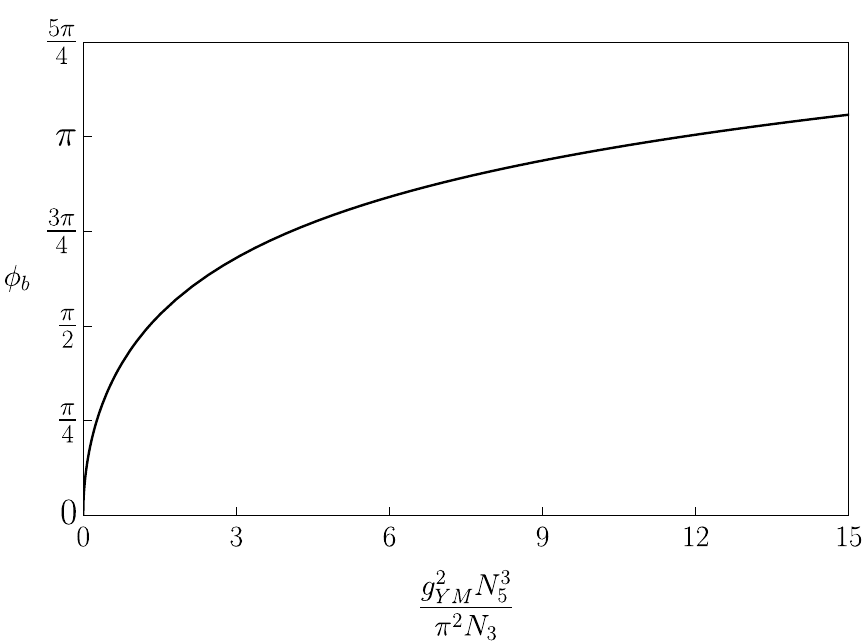}
    \caption{\label{D3D5BCFTPhibFig} The light crossing time of the D3/D5 BCFT as a function of field theory parameters.}
\end{figure}

We plot $\phi_{b}$ as a function of field theory parameters in Fig.~\ref{D3D5BCFTPhibFig}. While $\phi_b$ is always lower bounded by $\pi$ for the D3/D5 DCFT, for the BCFT it lives in the domain $\phi_b>0$. Indeed, since $\phi_{b} < \pi$ is allowed, this already implies that an ETW brane bottom-up model seemingly has greater potential to mimic the top-down construction. For sufficiently small $N_{5}$ compared to $N_{3}$, the AdS$_{4} \times S^{2} \times S^{2} \times \Sigma$ geometry will approximate to an effective AdS$_{5}$ geometry cut-off by an effective ETW brane who's location is given by the end of the strip. The notion that this really does behave like the bottom-up model will be further supported by the holographic entropy comparisons in the next subsection, but it should be noted that this prediction was first made in \cite{Coccia:2021lpp} (and further explored in \cite{Karch:2022rvr}).
 
\subsubsection{M2/M5 BCFT}\label{M2M5Phibsec}

Our other example is the background dual to the worldvolume theory of $N_2$ coincident M2 branes ending on $N_5$ coincident M5 branes. This worldvolume theory should be understood as ABJM theory on half-space at level $k=1$ with a suitable superconformal boundary condition preserving eight supercharges of SUSY. This boundary condition is presently unknown despite some effort~\cite{Hosomichi:2014rqa}. The dual geometry, a solution of 11d SUGRA, however is known~\cite{Bachas:2013vza} and was usefully post-processed in~\cite{Bena:2024dre}. The 11d spacetime is locally AdS$_3\times\mathbb{S}^3\times\mathbb{S}^3\times \Sigma$ with $\Sigma$ the upper half plane with a metric
\begin{equation}\label{M2M5Metric}
   ds^{2}_{11} = e^{2A}\left(f_{4}^{2}ds^{2}_{\text{AdS}_{3}} + f^2_{3}(d x^{2} + d y^{2}) + f_{1}^{2}ds^{2}_{\mathbb{S}^{3}} + f_{2}^{2} ds^{2}_{\bar{\mathbb{S}}^{3}}  \right)\,.
\end{equation}
The various warpfactors are
\begin{align}
\begin{split}
\label{M2M5warp}
    e^{2A}& = \left(\frac{h^{2} W_{+}W_{-}}{f_{1}^{2}} \right)^{1/3}
    \\
   f_{4}^{2} = \frac{1}{4(H \overline{H} - 1)} \,, &\qquad f_{3}^6 = \frac{1}{h^{2}} \,,\qquad  
    f_{1}^{2} = \frac{1}{W_{+}} \,, \qquad f_{2}^{2}= \frac{1}{W_{-}} \,,
\end{split}
\end{align}
where
\begin{equation}
    W_{\pm} = |H \pm i|^{2} + (H \overline{H} -1).
\end{equation}
and, in terms of a complex coordinate $w=x+iy$ on the upper-half plane ($y>0$) 
\begin{align}
\begin{split}
    h&= -i (w - \bar{w}) \,,\\
    H &= -\left(i \frac{w }{|w |} + \frac{\zeta \text{Im}(w)}{(\bar{w}-\xi )|w-\xi|} \right)\,.
\end{split}
\end{align}
The function $h$ is harmonic on the upper half plane, while $H$ satisfies an equation of Monge-Amp\'ere type. It is a superposition of two terms. The first ensures that $H$ obeys a certain boundary condition, $H = -\text{sgn}(x) i$ at $y=0$. The second corresponds to the stack of M5 branes with $\zeta$ and $\xi$ real parameters with $\zeta \xi >0$ that determine $N_2$ and $N_5$. The positivity of the product follows from the regularity of the 11d geometry. The geometry is weakly curved relative to the Planck scale if $\zeta/\ell_p^3$ and $\xi/\ell_p^3$ are both very large with $\zeta/\xi = \mathcal{O}(1)$.

This geometry caps off smoothly as $w\to 0$ and has an asymptotically AdS$_4\times\mathbb{S}^7$ region reached as $|w|\to\infty$ of the form
\begin{equation}
\label{E:M2M5asym}
	ds^2 = \frac{L^2}{4}ds^2_{\text{AdS}_4}+L^2 ds^2_{\mathbb{S}^7}\,, \qquad L^6 = 16 (\zeta^2+2\xi \zeta) \,.
\end{equation}
This radius is related to the number of M2 branes as
\begin{equation}
	\frac{L^6}{\ell_p^6} = 32\pi^2 N_2\,.
\end{equation}
Meanwhile, the M5 brane charge is calculated
\begin{equation}
	N_5 = \frac{\zeta}{\pi \ell_p^3}\,.
\end{equation}
These expressions allow us to trade $(\zeta,\xi)$ for field theory parameters as
\begin{align}
\begin{split}
	\zeta & = \pi N_5\ell_p^3\,,
	\\
	\xi &= \frac{\pi}{N_5}\left( N_2 - \frac{N_5^2}{2}\right)\ell_p^3\,.
\end{split}
\end{align}
A weakly curved background amounts to $N_2\gg 1$ and $N_5\gg 1$ with $N_2/N_5^2$ of $\mathcal{O}(1)$.

For positive $N_5$, i.e. positive $\zeta$ the condition $\xi\zeta >0$ then implies that $N_2$ is lower bounded as
\begin{equation}\label{M2M5lowerbound}
	N_2 \geq \frac{N_5^2}{2}\,.
\end{equation}
This result is a striking consequence of the regularity of the 11d background and its meaning in the dual BCFT is unknown. Taken literally, it suggests that these boundary conditions only exist for large enough $N_2$. If true it is tempting to speculate that the finite $N$ version of the bound is $N_2 \geq \frac{N_5(N_5+1)}{2}$, meaning that $N_2$ must be large enough to fill a triangular Young tableaux whose sides are of length $N_5$. Perhaps instead this bound is fictitious and in the bulk there is a geometric transition to another background when dialing $N_2$ below it.\footnote{We thank Michael Gutperle and Iosef Bena for discussions on these points.}

\begin{figure}[t]
    \centering
    \includegraphics[width=3.5in]{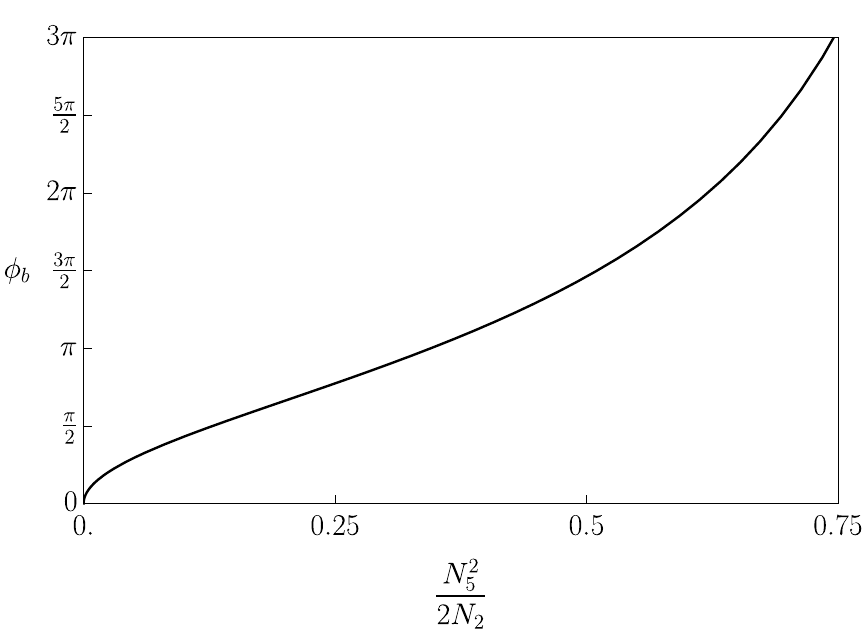}
    \caption{\label{M2M5BCFTPhibFig} The light crossing time of the M2/M5 BCFT as a function of field theory parameters.}
\end{figure}

Now we find the light crossing time. It is useful to work in polar coordinates $x = r \cos(\theta), y = r \sin(\theta)$ with $\theta \in [0, \pi]$. In these coordinates, our candidate null geodesic from the AdS boundary to the cap at $w=0$ sits at a point in the two $\mathbb{S}^3$'s and at $\theta=\pi$. (There is another candidate null geodesic at $\theta=0$ but this one has infinite time of flight.) The light crossing integrand for this geodesic is given by
\begin{equation}
    e^{-A(\rho)} \, d \rho = \frac{f_{3}}{f_{4}} dr =\frac{1}{(r+\xi)^2} \sqrt{ \zeta^2 + 2 \xi \zeta\left( 1 + \frac{\xi}{r}\right)} \, dr\,,
\end{equation}
which we integrate and find
\begin{equation}\label{phibunrefined}
     \phi_{b} = \frac{\zeta}{\xi} \sqrt{1 + \frac{2\xi}{\zeta} } + 2 \, \text{arccoth} \sqrt{1+\frac{2 \xi}{\zeta}} \,.
\end{equation}
Writing $\sigma = \frac{N_5^2}{2N_2} \in [0,1]$ we express the light crossing time as a function of field theory parameters as
\begin{equation}
\label{E:phiM2M5b}
	\phi_b = 2\left( \frac{\sqrt{\sigma}}{1-\sigma} + \text{arctanh}(\sqrt{\sigma})\right)\,,
\end{equation}
which we plot in Fig.~\ref{M2M5BCFTPhibFig}. Similarly to the D3/D5 BCFT case, the light crossing time exists over the whole range $\phi_{b} > 0$, and not only $\phi_{b} \geq \pi$ as for our top-down defects.

\subsection{Boundary entropy and comparing with Takayanagi's model}

As in Subsection~\ref{DefectComps} we compare the boundary entropy $B_0$ of these top-down BCFTs with that obtained in Takayanagi's model of holographic BCFT. The boundary entropy of the D3/D5 BCFT has been computed in~\cite{Estes:2014hka}. We review it in the Appendix. The entropy of the M2/M5 BCFT has not been computed before; we find it below. With these results in hand we compare.

\subsubsection{M2/M5 boundary entropy and boundary central charge}\label{S:M2M5b}

As we mentioned in Subsection~\ref{ETWtoKR} there are many possible notions of boundary entropy in $d>2$ dimensional BCFT. Ours, following~\cite{Nozaki:2012qd,Estes:2014hka,Kobayashi:2018lil}, is through the vacuum entanglement entropy across a hemisphere of radius $R$ centered on the boundary.  See Eqs.~\eqref{E:entropy1} and~\eqref{E:entropy2}. 

In AdS/CFT this entanglement entropy is calculated through the Ryu-Takayanagi formula in terms of the area $\mathcal{A}$ of a codimension-2 minimal surface at fixed time anchored on the conformal boundary at the entangling surface through
\begin{equation}
	S_{\rm EE}  = \frac{\mathcal{A}}{4G}\,,
\end{equation}
with $G$ the 11d Newton's constant. Here $16\pi G = (2\pi)^8 \ell_p^9$. The 11d background~\eqref{M2M5Metric} has the form
\begin{equation}
\label{E:generalBCFTmetric}
	ds^2 = f(x,y^a)^2 ds^2_{\text{AdS}_3} + \rho(x,y^a)^2 dx^2 + G_{ab}(x,y^c) dy^ady^b\,,
\end{equation}
and has an asymptotically AdS$_4\times \mathbb{S}^7$ region at large positive $x$. The minimal surface in such a background has been found in~\cite{Jensen:2013lxa}. Writing
\begin{equation}
	ds^2_{\text{AdS}_3} = \frac{-dt^2 + dz^2 + du^2}{u^2}\,,
\end{equation}
it is simply a surface that is extended in $(x,y^a)$ and localized to
\begin{equation}
	u^2 + z^2 = R^2\,.
\end{equation}
Using the results of~\cite{Estes:2014hka} the entanglement entropy can then be written as
\begin{align}
\begin{split}
\label{sphericalEntropy}
    S_{\rm EE} &= \frac{R}{2 G} \int dy^{a} dx du \, \sqrt{\det G} \frac{\rho f}{u\sqrt{R^2-u^2}} 
    \\
    &= \frac{(2\pi^2)^2 R}{2G}\int_0^{\pi} d\theta \int_{\varepsilon u_c(\theta)}^R du\int_0^{r_c\left(\frac{\varepsilon}{u},\theta\right)}dr   \frac{r e^{9A}f_1^3f_2^3f_3^2f_4}{u\sqrt{R^2-u^2}} \,.
\end{split}
\end{align}
In going from the first line to the second we have matched the warpfactors in~\eqref{E:generalBCFTmetric} to those of~\eqref{M2M5Metric}, identified the variable $x$ with the radial coordinate $r=|w|$, and integrated over the $\mathbb{S}^3$'s. This integration is performed with respect to certain cutoffs in $(r,u)$ that are described in detail in\cite{Estes:2014hka}. In field theory terms they encode a short-distance regulator $\varepsilon\ll 1$ appearing in~\eqref{E:entropy2}, which in the gravity computation regulates the infinite area near the AdS boundary. The integrand is
\begin{equation}
	\frac{r e^{9A}f_1^3f_2^3f_3^2f_4}{u\sqrt{R^2-u^2}} = \frac{r^4\sin^3(\theta)\left( \zeta^2 + 2\zeta \xi \frac{\chi}{r}\right)}{u \chi^4\sqrt{R^2-u^2}}\,, \qquad \chi = |w-\xi| = \sqrt{r^2+\xi^2-2r\xi\cos(\theta)}\,.
\end{equation}
To find the cutoff $r_c$ we must specify the asymptotic form of the metric~\eqref{E:generalBCFTmetric}. At large $r$ it has the form
\begin{align}
\begin{split}
    f^{2} &= \frac{L^2}{4} \left(r^2 e^{2c} + f^{(-1)}(y) r+\mathcal{O}(1) \right)\,, 
    \\
    \rho^{2} &= \frac{L^2}{r^2} \left(1 + \rho^{(1)}(y)\frac{1}{r} +\mathcal{O}(r^{-2}) \right)\,, 
    \\
    G_{ab}dy^a dy^b &= \left( G^{(0)}(y)_{ab}+ G^{(1)}(y)_{ab}\frac{1}{r} + \mathcal{O}(r^{-2})\right)dy^ady^b\,,
\end{split}
\end{align}
where $L$ is the asymptotic AdS radius given in~\eqref{E:M2M5asym}, which we recapitulate here along with giving the constant $c$,
\begin{equation}
	L^6 = 16 (\zeta^2 + 2\zeta \xi)\,, \qquad e^{2c} = \frac{4}{\zeta^2+2\zeta \xi}\,,
\end{equation}
and the other quantities are functions of the $y^a$. In particular,
\begin{align}
\begin{split}
	f^{(-1)}(y) & =  - \frac{8(\zeta^2+6\zeta \xi + 6 \xi^2)}{3\zeta(\zeta+2\xi)^2}\cos(\theta)\,,
	\\
	\rho^{(-1)}(y) & = - \frac{2(\zeta^2-3\xi^2)}{3(\zeta+2\xi)}\cos(\theta)\,,
	\\
	G^{(0)}_{ab}dy^ady^b &= L^2 \left( d\left( \frac{\theta}{2}\right)^2 + \sin^2\left( \frac{\theta}{2}\right)ds^2_{\mathbb{S}^3} + \cos^2\left( \frac{\theta}{2}\right)ds^2_{\mathbb{\bar{S}}^3}\right)\,.
\end{split}
\end{align}
By~\cite{Estes:2014hka} the large $r$ cutoff surface is located at 
\begin{equation}
	\ln r_c = \ln \left( \frac{2 u}{\varepsilon}\right) -c + \frac{e^c \rho^{(-1)}(\theta)}{4}\frac{\varepsilon}{u} + \mathcal{O}\left( \frac{\varepsilon^2}{u^2}\right)\,.
\end{equation}
The cutoff surface $u = \varepsilon u_c(\theta)$ is more or less arbitrary for $\mathcal{O}(1)$ functions $u_c(\theta)$. 

Picking $u_c$ to be a constant we can perform the $r$-integral, then the $\theta$-integral, then the $u$-integral. The $r$-integral, and its $\theta$-integral can be performed in closed form with rather complicated expressions appearing at intermediate steps. The final result is
\begin{align}
\begin{split}
	S_{\rm EE}& = \int_{\varepsilon u_c}^R du \frac{(2\pi^2)^2R}{2G}\int_0^{\pi} d\theta \int_0^{r_c\left( \frac{\varepsilon}{u},\theta\right)} dr \frac{r e^{9A}f_1^3f_2^3f_3^2f_4}{u\sqrt{R^2-u^2}} 
	\\
	& = \frac{R}{8\pi^3 \ell_p^9} \int_{\varepsilon u_c}^R \frac{du}{u\sqrt{R^2-u^2}}\left( \frac{32L^9 u}{\varepsilon} -2 \zeta \xi^2  + O\left( \frac{\varepsilon}{u}\right)\right)\,.
\end{split}
\end{align}
The first term in the second line contributes to a perimeter term $\propto \frac{R}{\varepsilon}$ in the EE, the second to the logarithmic term encoding the boundary entropy $B_0$, and the neglected terms contribute to the constant in the EE. That is,
\begin{equation}
	S_{\rm EE} = \frac{a_1}{\varepsilon} + B_0 \ln \left( \frac{2R}{\varepsilon}\right) + \mathcal{O}(1)\,,
\end{equation}
with
\begin{equation}
\label{E:M2M5bdyEntropy}
	B_0 = -\frac{\zeta \xi^2}{4\pi^3 \ell_p^9} = - \frac{\left( N_2 - \frac{N_5^2}{2}\right)^2}{4N_5}\,.
\end{equation}

Since the 11d SUGRA background is weakly curved for $N_2,N_5\gg 1$ with $N_2/N_5^2$ held fixed, we note that $B_0$ is of $\mathcal{O}(N_2^{3/2})$ as one would expect for a boundary condition in the ABJM theory at $k=1$. Thinking of the background instead as dual to a self-dual string in the $\mathcal{N}=(2,0)$ theory, we see that it has the expected scaling of $\mathcal{O}(N_5^3)$.

The boundary entropy $B_0$ is determined by the $b$-type central charge of the dual BCFT. It is $b = 3 B_0$~\cite{Fursaev:2016inw,Jensen:2018rxu} giving
\begin{equation}
\label{E:bM2M5}
	b  = - \frac{3}{4N_5} \left( N_2 - \frac{N_5^2}{2}\right)^2\,.
\end{equation}
This quantity is known to decrease under boundary RG flow~\cite{Jensen:2015swa} in a result called the ``$b$-theorem.'' We note that the value of $b$ in~\eqref{E:bM2M5} is consistent with the $b$-theorem: by slightly separating some number $\Delta N_5$ of the M5 branes from the M2 branes in their mutually transverse directions we can arrange for the gravity dual of a boundary RG flow starting from the BCFT described by $N_2$ M2 branes ending on $N_5$ M5 branes, and ending at the one where they end on $N_5'=N_5-\Delta N_5 < N_5$ M5 branes. Since for $N_2 > \frac{N_5^2}{2}$
\begin{equation}
	-\left.\frac{db}{dN_5}\right|_{N_2} = -\frac{3N_2}{4}\left( \frac{N_2}{N_5^2} + 1 - \frac{4N_5^2}{3N_2}\right) \leq 0\,,
\end{equation}
so that $b$ strictly decreases with decreasing $N_5$, we have
\begin{equation}
	b_{\rm UV} - b_{\rm IR} \geq 0\,,
\end{equation}
in accordance with the $b$-theorem.

\subsubsection{Comparing}

\begin{figure}[t]
	\centering
	\begin{minipage}{0.45\textwidth}
 		\centering
		\includegraphics[width=3in]{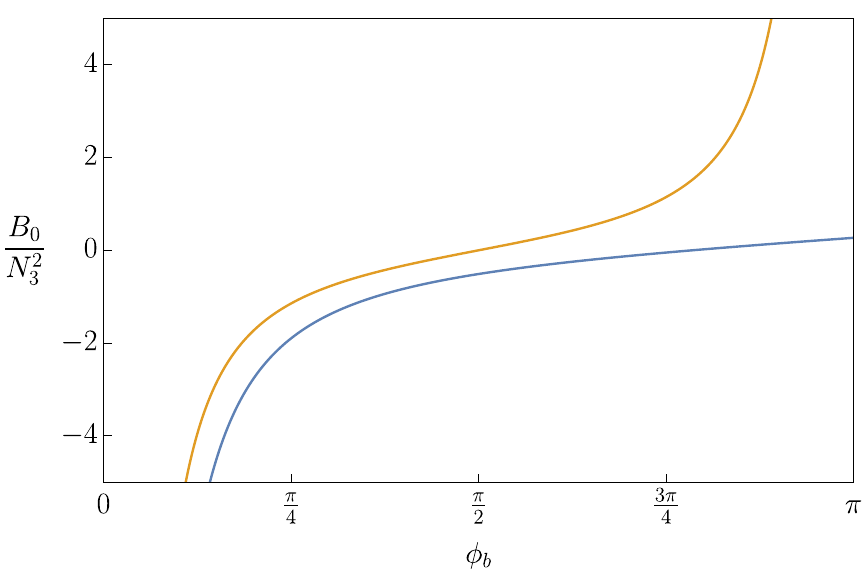}
		\caption{\label{d4BCFTCompFig} The normalized boundary entropy of the D3/D5 BCFT (blue), and that of Takayanagi's bottom-up model (orange), plotted as a function of light crossing time.}
	\end{minipage}
	\hfill
	\begin{minipage}{0.45\textwidth}
   		\centering
		\includegraphics[width=3in]{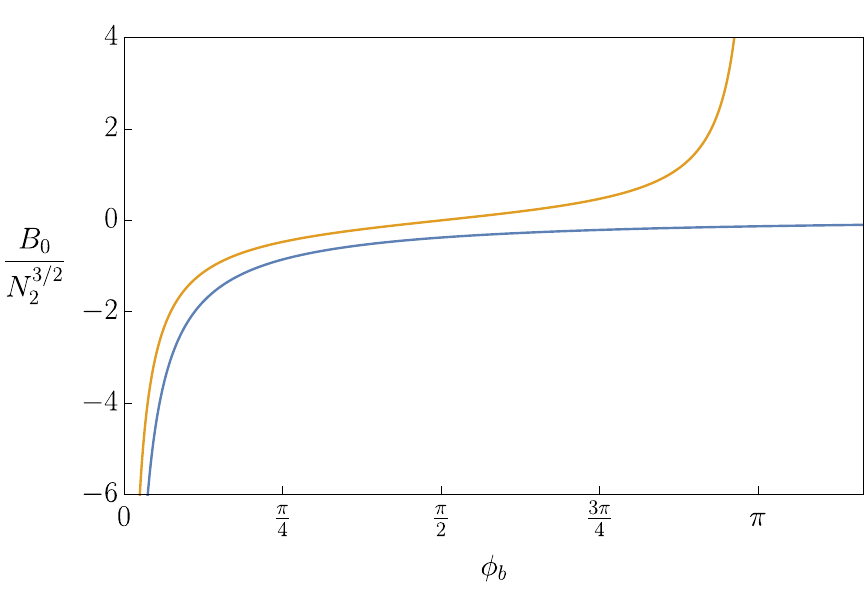}
		\caption{\label{d3BCFTCompFig} The normalized boundary entropy of the M2/M5 BCFT (blue), and that of Takayanagi's bottom-up model (orange), plotted as a function of light crossing time.}
	\end{minipage}
\end{figure}

As in Subsection~\ref{DefectComps} we now compare the boundary entropy of our top-down models with those of Takayanagi's bottom-up model. In each case we plot normalized versions of the boundary entropy, $B_0/N_3^2$ in $d=4$ and $B_0/N_2^{3/2}$ in $d=3$. However, we first need to trade the Newton's constant of the bottom-up model for an effective rank of a dual gauge group. In Subsection~\ref{DefectComps} we have already done this for $d=4$. In $d=3$ and using that the physical radius of the AdS$_4$ in the AdS$_4\times\mathbb{S}^7$ vacuum of M-theory is $\tilde{L} = \frac{L}{2}$ together with $\frac{L^6}{\ell_p^6} = 32\pi^2 N_2$ we have
\begin{equation}
\label{E:d3MatchingG}
	\frac{\tilde{L}^2 (2\tilde{L})^7 \text{vol}(\mathbb{S}^7)}{16\pi G_{11}} = \frac{\tilde{L}^9}{6\pi^4 \ell_p^9} = \frac{\tilde{L}^2}{16\pi G_4}\,,  \qquad \Rightarrow \qquad \frac{\tilde{L}^2}{G_4} = \frac{2^{3/2}}{3}N_2^{3/2}\,.
\end{equation}
We plot the normalized boundary entropies in Figs.~\ref{d4BCFTCompFig} and~\ref{d3BCFTCompFig}.

There are some qualitative features that are obvious from both plots. First, all boundary entropies monotonically increase with $\phi_b$ from negative infinity. Second, unlike top-down DCFTs, there are top-down BCFTs with $\phi_b <\pi$ and in particular $\phi_b<\frac{\pi}{2}$ which, when realized with bottom-up branes, correspond to ETW branes of negative tension. Third, unlike the bottom-up examples, $\phi_b$ is unbounded from above in top-down constructions. Fourth, the boundary entropy of our M2/M5 example is upper-bounded by zero rather than positive infinity as in the bottom-up example. Lastly, at least in the range of smaller $\phi_b \sim (0,\pi/2)$, the agreement between top-down and bottom-up constructions is surprisingly good. 

Both the bottom-up and top-down models have large negative boundary entropy as $\phi_b\to 0$. We wrap up with a perturbative comparison in that regime. Using Eqs.~\eqref{universalbottomupentropy} (along with the relation $\tanh(\rho_0) = \cos(\phi_b)$ for the bottom-up models),~\eqref{E:d4MatchingG}, and~\eqref{E:d3MatchingG}, we obtain the asymptotic form of the boundary entropy of the bottom-up models at small $\phi_b$. We do the same in our top-down examples using Eqs.~\eqref{phibD3D5BCFT},~\eqref{E:phiM2M5b},~\eqref{E:D3D5B}, and~\eqref{E:M2M5bdyEntropy}. The $d=4$ results are
\begin{equation}
	\frac{B_0}{N_3^2} = \begin{cases}-\frac{1}{2\phi_b^2} + \frac{1}{2}\ln(\phi_b) + \mathcal{O}(\phi_b^0) \,, & \text{bottom-up}\,, \\ -\frac{\pi^2}{12\phi_b^2} + \frac{1}{2}\ln(\phi_b)+\mathcal{O}(\phi_b^0)\,, & \text{D3/D5} \,,\end{cases}
\end{equation}
while in $d=3$ we have
\begin{equation}
	\frac{B_0}{N_2^{3/2}} = \begin{cases} -\frac{2}{3\sqrt{2}\phi_b} + \mathcal{O}(\phi_b)\,, & \text{bottom-up}\,, \\ - \frac{1}{\sqrt{2}\phi_b} + \mathcal{O}(\phi_b)\,,  & \text{M2/M5}\,.\end{cases}
\end{equation}

\section{Discussion}\label{discussion}

We now summarize our main findings. We begin by reiterating the observation of~\cite{Reeves:2021sab} that holographic BCFTs and DCFTs are special in that they have a universal approximate singularity in correlation functions. This singularity is governed by the quantity $\phi_b$ that can be interpreted as the amount of global boundary time it takes for a light ray to travel from the boundary to the end-of-the-world (for a BCFT) or to the other part of the boundary (for a DCFT). This light crossing time plays a starring role in our analysis.

In the bottom-up models with a tensionful domain wall or ETW brane the light-crossing time is a monotonically increasing function of brane tension. As such we can characterize bottom-up models by their tension, or by their light crossing time. In the Karch-Randall model for DCFT $\phi_b$ occupies the range $[0,2\pi)$, with $\pi$ corresponding to a tensionless KR brane, i.e. empty AdS, negative tension to the range $[0,\pi)$, and positive tension to the range $(\pi,2\pi)$. In Takayanagi's model for BCFT it behaves in the same way as for the Karch-Randall model, but with $\phi_b \to \phi_b/2$, e.g. a tensionless ETW brane corresponds to $\phi_b=\pi/2$. 

We computed $\phi_b$ for several top-down DCFTs and BCFTs. For DCFTs we always found $\phi_b\geq \pi$; in the Appendix we further demonstrate that linearized Janus perturbations, even for relevant operators, always lead to $\phi_b\geq \pi$. In all of our examples we can achieve $\phi_b >2\pi$, and moreover the light crossing time is a well-defined function of field theory parameters. The basic question we have in mind in this work is the extent to which bottom-up models can imitate top-down constructions. In the context of defects, only those defects for which $\pi \leq \phi_b<2\pi$ can be imitated (whether well, or poorly) by a KR brane, which moreover has positive tension. Those with $\phi_b>2\pi$ cannot be imitated, and there is no top-down version of a negative tension KR brane.

It seems reasonable to conjecture that holographic DCFTs always obey the bound $\phi_b \geq \pi$. If this bound is true, then we suspect it relies in part on unitarity for the following reason. Consider non-SUSY Janus deformations where we denote the massless bulk scalar field as $\varphi$. Taking the flux of $\varphi$ to be imaginary rather than real, even perturbatively in the deformation, leads to $\phi_b <\pi$. Such a deformation violates unitarity from the point of view of the dual CFT, and indeed a similar construction was recently used to generate traversable wormholes connecting two asymptotically AdS boundaries~\cite{Kawamoto:2025oko} without directly correlating the two boundaries as in~\cite{Gao:2016bin,Maldacena:2018lmt,Harvey:2023oom}.

In our two top-down BCFT examples we find that $\phi_b$ is a well-defined function of field theory parameters, and that by adjusting those we can achieve any $\phi_b >0$. In particular there is no non-trivial lower bound on $\phi_b$. There are top-down BCFTs with $\phi_b >\pi$ that cannot be imitated by a tensionful ETW brane, and BCFTs that can be imitated by ETW branes with negative tension.

We proceed to compare the physics of bottom-up and top-down models, examining the physics of both as a function of $\phi_b$. We focus on defect/boundary entropy. Those entropies are monotonically increasing functions of $\phi_b$ both in bottom-up and top-down models. Our comparisons for DCFTs may be found in Figs.~\ref{d4DCFTCompFig} and~\ref{d2DCFTCompFig}, while those for BCFTs are in Figs.~\ref{d4BCFTCompFig} and~\ref{d3BCFTCompFig}. Our main conclusion is that the bottom-up models qualitatively approximate the behavior of the top-down ones. In the regime of BCFTs with $0<\phi_b<\pi/2$, effectively negative tension ETW branes, the two match remarkably well, although unlike in the bottom-up model, nothing special happens in top-down constructions at the value $\phi_b = \pi/2$ where the bottom-up entropy changes sign.

Along the way we computed the boundary central charge $b$~\cite{Jensen:2015swa} of the M2/M5 BCFT, the worldvolume theory of $N_2$ M2 branes ending on $N_5$ coincident M5 branes. The result is given in~\eqref{E:bM2M5}. We also noticed a curious property of the supergravity background dual to the BCFT~\cite{Bachas:2013vza} suggesting that $N_2$ is lower-bounded as $N_2 \geq \frac{N_5^2}{2}$. It is unclear if such a bound is genuine or if instead BCFTs below the bound exist but the bulk undergoes a geometric transition to a different background. In any case our result for $b$ is consistent with the $b$-theorem of~\cite{Jensen:2015swa}, which in this setting is the statement that $b$ decreases under boundary RG flow. 

While the 11d supergravity background dual to the M2/M5 BCFT is known (although the dual boundary condition on ABJM theory is not), the background dual to the localized M2/M5 DCFT is presently unknown. The authors of~\cite{Bachas:2013vza} have already done most of the work of finding it, reducing the problem to finding a certain complex function on the strip obeying a linear first-order differential equation of Monge-Amp\'ere type obeying known boundary conditions. Solving that problem would be independently interesting and useful.

\subsection*{Acknowledgements}

It is a pleasure to thank I.~Bena, M.~Gutperle, S.~P.~Harris, A.~Karch, D.~Mazac, D.~Neuenfeld, M.~Rozali, S.~Zhibodoev, and C.~Waddell for useful discussions. This work was supported in part by an NSERC Discovery Grant.

\appendix

\section{Light crossing time in Einstein-scalar models}\label{ETWcalcs}

In this Appendix we compute the light crossing time in three classes of models all of which involve Einstein gravity with negative cosmological constant coupled to a scalar field. The first has a bottom-up tensionful ETW brane and we study the response of the light crossing time to linear perturbations of the scalar field. In the second we study the response around that of empty AdS to linearized Janus deformations. Lastly we consider Janus-like interfaces in which the dual theory has gravity coupled to a massless scalar. The non-SUSY Janus solution discussed in Subsection~\ref{nonSUSYJanusSec} is an example of the latter. In the first model scalar perturbations tend to increase $\phi_b$, and in the later two we always find $\phi_b\geq \pi$.

\subsection{ETW brane coupled to a scalar field}

In this Appendix we study the model
\begin{align}
\begin{split}
\label{E:perturbedETWaction}
    S &= \frac{1}{16\pi G}\int_N d^{d+1}x \sqrt{-g}(R-2\Lambda) + \frac{1}{8\pi G}\int_Q d^dx \sqrt{-h}(K-T) 
    \\
    & \qquad \qquad - \frac{1}{2}\int_N d^{d+1}x \sqrt{-g}((\partial\varphi)^2 + m^2\varphi^2) + (\text{terms on }M)\,.
\end{split}
\end{align}
Here $N$ is the bulk spacetime, $Q$ the ETW brane, and $M$ the conformal boundary. We take $\Lambda = - \frac{d(d-1)}{2}$ so that the vacuum Einstein's equations admit AdS$_{d+1}$ solutions of unit radius. Treating this model as the dual of a BCFT, the vacuum state in the absence of a scalar background is a cutout of AdS$_{d+1}$ as discussed in Section~\ref{S:ETW}. The geometry is
\begin{equation}
    ds^2 = \cosh^2(\rho) ds^2_{\text{AdS}_d}+d\rho^2\,,
\end{equation}
for $\rho \geq \rho_0$ with $\tanh(\rho_0) = - \frac{T}{d-1} = \cos(\phi_b)$ with $\phi_b\in [0,\pi]$. Now we turn out a scalar background in the linearized approximation in such a way that is consistent with the symmetries of a BCFT and compute the perturbation in $\phi_b$. Usually, but not always, we find that a scalar background increases $\phi_b$.

\subsubsection{Two contributions}

We will study perturbative solutions to the equations of motion that follow from~\eqref{E:perturbedETWaction}. The most general metric and scalar background consistent with the symmetries of a dual BCFT are
\begin{align}
    \begin{split}
    \label{E:BCFTbackground}
        \varphi &= \varphi(\rho)\,,
        \\
        ds^2 &= e^{2A(\rho)}ds^2_{\text{AdS}_{d}}+d\rho^2\,.
    \end{split}
\end{align}
Perturbing around $\varphi = 0$ and the geometry a cutout of AdS we take for $\epsilon\ll 1$ a formal expansion parameter
\begin{align}
    \begin{split}
\label{E:perturbExpansion}
    \varphi (\rho)& =  \epsilon \, \varphi^{(1)} (\rho) + \mathcal{O}(\epsilon^{2})\,,
    \\
    A(\rho) & = \ln \cosh(\rho) + \epsilon^2 A^{(2)}(\rho) + \mathcal{O}(\epsilon^4)\,.
\end{split}
\end{align}
The location of the ETW brane, $\rho = \rho_0$, is also corrected as
\begin{equation}
    \rho_0 = \rho_0^{(0)} + \epsilon^2 \rho_0^{(2)}+\mathcal{O}(\epsilon^4)\,,
\end{equation}
where $\tanh(\rho_0^{(0)}) = - \frac{T}{d-1} $. This
in turn leads to an expansion of the light crossing time
\begin{equation}
    \phi_{b} =\int_{\rho_0}^{\infty} d\rho \,e^{-A(\rho)} =  \phi_{b}^{(0)} + \epsilon^{2} \, \phi_{b}^{(2)} + \mathcal{O}(\epsilon^{4})\,,
\end{equation}
where $\tanh(\rho_0^{(0)}) = \cos(\phi_b^{(0)})$ and
\begin{equation}
\label{E:perturbedPhib}
    \phi_b^{(2)} = -\int_{\rho_0^{(0)}}^{\infty} \frac{d\rho}{\cosh(\rho)}A^{(2)}(\rho) - \frac{\rho_0^{(2)}}{\cosh(\rho_0^{(0)})}\,.
\end{equation}
So the perturbation $\phi_b^{(2)}$ in the light crossing time receives two distinct contributions, one coming from the perturbation of the warpfactor and the other from the perturbation of the location of the ETW brane. 

\subsubsection{Computing $\phi_{b}^{(2)}$}

The analogue of the Israel junction condition
\begin{equation}
    A'(\rho_0) = - \frac{T}{d-1}\,,
\end{equation}
determines the location of the ETW brane. Expanding it to second order in $\epsilon$ gives the condition
\begin{equation}
\label{E:perturbedETW}
    \rho_0^{(2)} = - A'^{(2)}(\rho_0^{(0)}) \cosh^2(\rho_0^{(0)})\,,
\end{equation}
which determines the perturbation $\rho_0^{(2)}$ once we have the perturbation $A^{(2)}$ in the warpfactor. The latter is determined by the linearized Einstein-scalar equations.

Our strategy for the latter is to first solve the Klein-Gordon equation for the scalar at $\mathcal{O}(\epsilon)$, then to solve the Einstein's equations at $\mathcal{O}(\epsilon^2)$ in which the $\mathcal{O}(\epsilon)$ scalar profile contributes an $\mathcal{O}(\epsilon^2)$ source. The Klein-Gordon equation is
\begin{equation}
    \left( -\Box + m^2\right)\varphi = \epsilon\left( - \partial_{\rho}^2 - d \tanh(\rho) \partial_{\rho} + \Delta(\Delta - d)\right)\varphi^{(1)} + \mathcal{O}(\epsilon^3)=0\,,
\end{equation}
where we have used $m^2 = \Delta(\Delta-d)$ with $\Delta$ the conformal dimension of the operator dual to the scalar field $\varphi$. Solving the $\mathcal{O}(\epsilon)$ term gives
\begin{equation}
\label{E:phi1}
    \varphi^{(1)} = \text{sech}^{d/2}(\rho)\left( c_1 P_{\frac{d}{2}-1}^{\frac{d}{2}-\Delta}(\tanh(\rho)) + c_2 Q_{\frac{d}{2}-1}^{\frac{d}{2}-\Delta}(\tanh(\rho))\right)\,,
\end{equation}
for some constants $c_1$ and $c_2$. At large $\rho$ this solution behaves as
\begin{equation}
    \varphi^{(1)} \approx C_+ e^{(\Delta -d)\rho} + C_- e^{-\Delta \rho}\,,
\end{equation}
with $C_+\propto c_2$ and $C_-$ a linear combination of $c_1$ and $c_2$. The term $C_+$ corresponds to a position-dependent source $J\propto \frac{C_+}{(x_{\perp})^{d-\Delta}}$, which is consistent with the residual conformal symmetry. When $\Delta<d$ such a source is allowed in a renormalizable theory, but for $\Delta>d$ it is not. So in the former case we allow for both $c_1$ and $c_2$, while in the latter only $c_1$ may be nonzero. The other boundary condition we impose is at the ETW brane. Instead of imposing a Neumanmn or Dirichlet boundary condition on $\varphi$ there we allow for any behavior, which can be made consistent with a variational principle for $\varphi$ provided we allow a $\varphi^2$ term in the ETW brane action.

Now for the Einstein's equations. It is convenient to solve the $\rho\rho$-component, 
\begin{equation}
    R_{\rho\rho} - \frac{R}{2}g_{\rho\rho} - \frac{d(d-1)}{2}g_{\rho\rho} = 8 \pi G T_{\rho\rho}\,,
\end{equation}
which implies the other Einstein's equations. Using $T_{\rho\rho} = \epsilon^2T^{(2)} + \mathcal{O}(\epsilon^4)$ with
\begin{equation}
\label{E:T2}
    T^{(2)} = \frac{1}{2}(\partial_{\rho}\varphi^{(1)})^2 - \frac{1}{2}\Delta(\Delta-d)(\varphi^{(1)})^2\,,
\end{equation}
it becomes
\begin{equation}
   \epsilon^2 \left[ d(d-1) \tanh^2(\rho) \partial_{\rho}\left( \frac{A^{(2)}}{\tanh(\rho)}\right) - 8\pi G T^{(2)}  \right] + \mathcal{O}(\epsilon^4) = 0\,,
\end{equation}
which has the solution
\begin{equation}
\label{E:A2}
    A^{(2)} = -\tanh(\rho)\int_{\rho}^{\infty} d\rho'\,S(\rho')\,, \qquad S(\rho) = \frac{8 \pi G}{d(d-1)}\coth^2(\rho)T^{(2)}\,.
\end{equation}
For $\rho^{(0)}_0>0$, i.e. a negative tension ETW brane, this expression is unambiguous. However for a positive tension ETW brane with $\rho_0^{(0)}<0$ it has to be regulated on account of the divergence in the integrand at $\rho=0$. We therefore define two regulated versions of the source $S(\rho)$,
\begin{align}
\begin{split}
\label{E:regularizedS}
    \widetilde{S}(\rho)& = S(\rho) - \frac{8\pi G}{d(d-1)}\frac{T^{(2)}(0)}{\rho^2}\,,
    \\
    \widetilde{S}_2(\rho)& = \frac{S(\rho)}{\cosh(\rho)} - \frac{8\pi G}{d(d-1)}\frac{T^{(2)}(0)}{\rho^2}\,.
\end{split}
\end{align}
The first gives a regulated version of the solution~\eqref{E:A2} for $A^{(2)}$
\begin{equation}
\label{E:newA2}
    A^{(2)}(\rho) = - \tanh(\rho)\left( \int_{\rho}^{\infty} d\rho' \,\widetilde{S}(\rho)    + \frac{8\pi G}{d(d-1)} \frac{T^{(2)}(0)}{\rho}\right)\,. 
\end{equation}
Using this, the regulated source $\widetilde{S}_2$, and the perturbation in the ETW location~\eqref{E:perturbedETW}, the perturbation~\eqref{E:perturbedPhib} $\phi_b^{(2)}$ of the light crossing time simplifies to become
\begin{align}
\begin{split}
    \phi_b^{(2)} =   -\int_{\rho_0^{(0)}}^{\infty}d\rho\,\widetilde{S}_2(\rho)      + \sinh(\rho_0^{(0)})S(\rho_0^{(0)}) - \frac{8\pi G}{d(d-1)}\frac{T^{(2)}(0)}{\rho_0^{(0)}}\,.
\end{split}
\end{align}

The source $S$ is given in~\eqref{E:A2}, the regulated source $\widetilde{S}_2$ in~\eqref{E:regularizedS}, $T^{(2)}$ in~\eqref{E:T2}, and the scalar profile $\varphi^{(1)}$ in~\eqref{E:phi1}. The perturbation $\phi_b^{(2)}$ is quadratic in the parameters $c_1,c_2$ governing the scalar profile. For $\Delta <d$ where they may both be turned on, we may write
\begin{equation}\label{compactphib}
    \phi_{b}^{(2)} = \vec{c}^{\, \, T} M \vec{c}\,, \qquad \vec{c} = \begin{pmatrix} c_1 \\ c_2 \end{pmatrix}\,,
\end{equation}
for a complicated matrix $M$ with eigenvalues $\lambda_{\pm}$ and $\lambda_+\geq \lambda_-$. The perturbation $\phi_b^{(2)}$ is positive when both eigenvalues are positive, and can be negative when one or both are negative. 

\begin{figure}[tp]
    \centering
    \begin{minipage}{.45\textwidth}
    \includegraphics[width=2.7in]{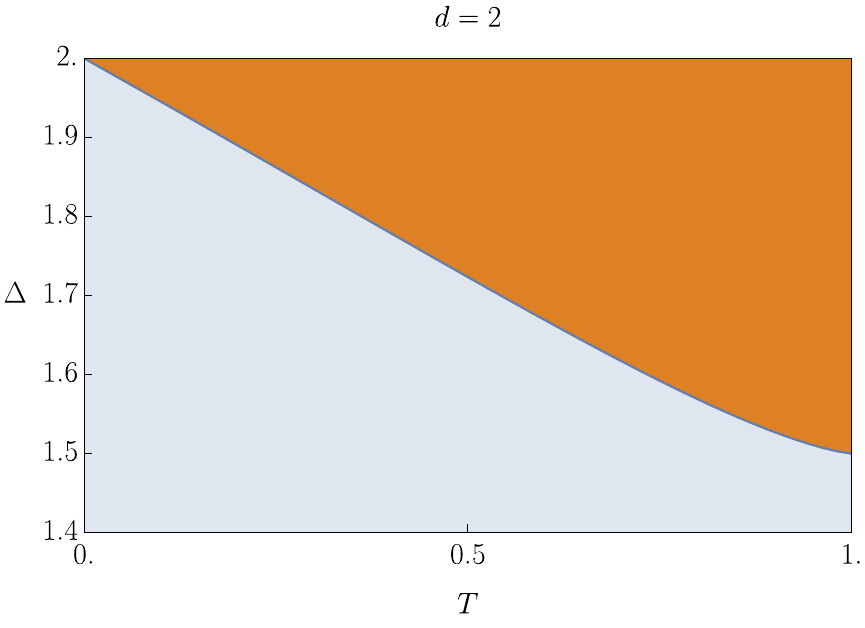}
    \end{minipage}
    \hfill
    \begin{minipage}{.45\textwidth}
    \includegraphics[width=2.7in]{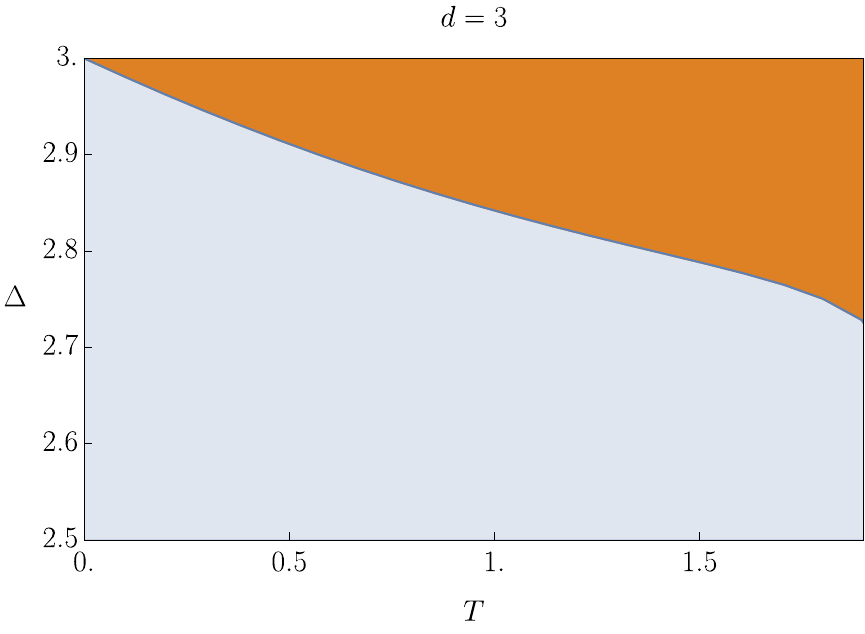}
    \end{minipage}
    \\
    \begin{minipage}{.45\textwidth}
    \includegraphics[width=2.7in]{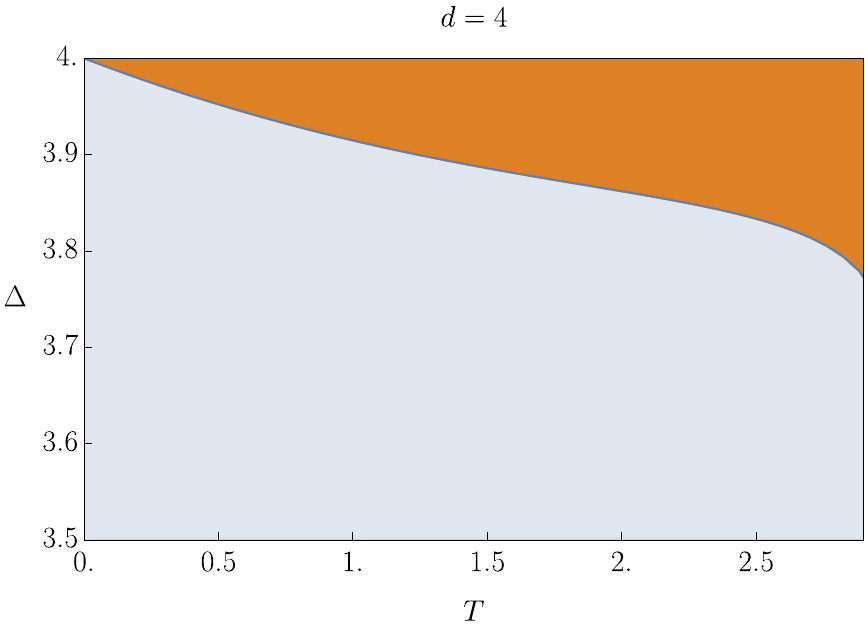}
    \end{minipage}
    \hfill
    \begin{minipage}{.45\textwidth}
    \includegraphics[width=2.7in]{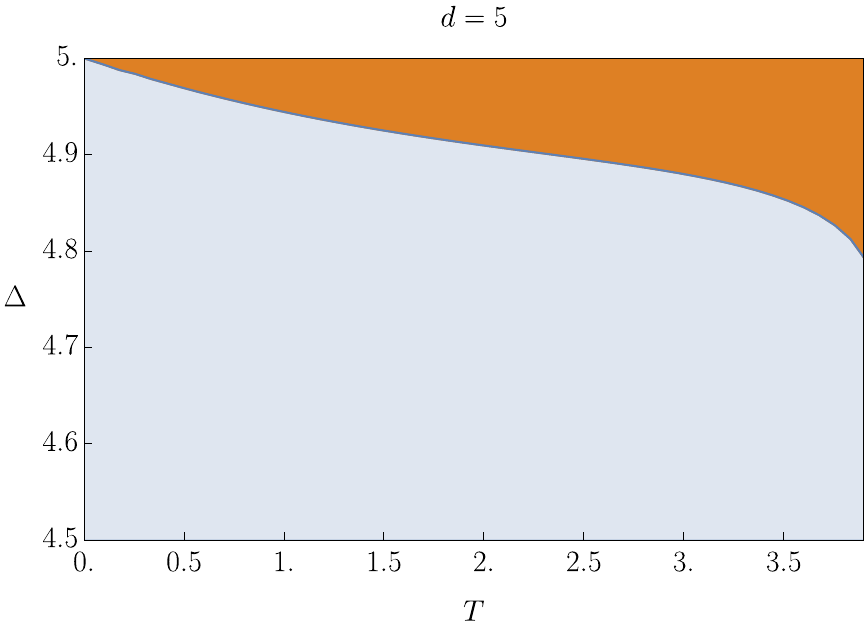}
    \end{minipage}
    \\
    \includegraphics[width=2.7in]{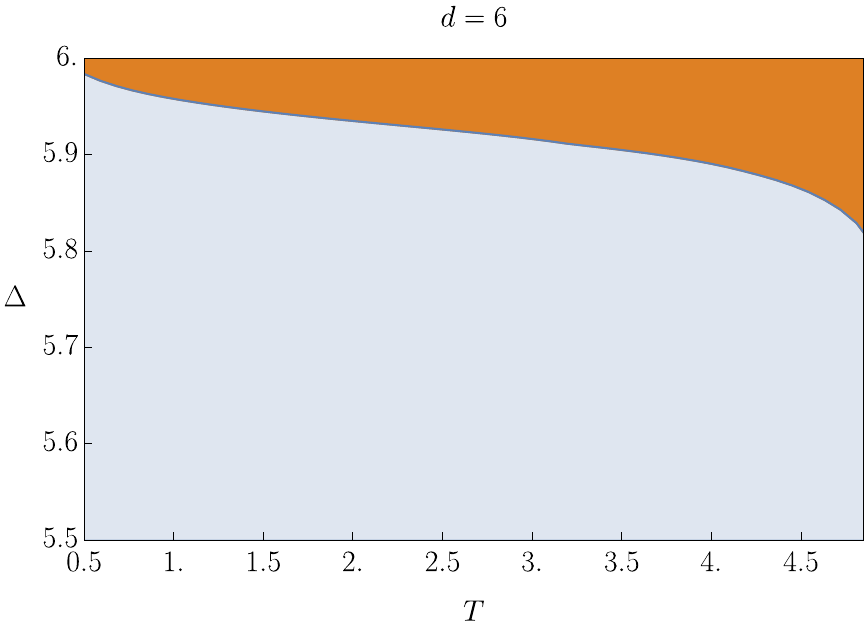}\,
    \caption{\label{LightCrossingChanges} The sign of the smallest eigenvalue $\lambda_-$ governing the change in the light crossing time $\phi_{b}^{(2)}$. For $d = 2$, we analytically solved \eqref{compactphib} to extract $\lambda_{-}$, and our numerical integrations for $d = 3,4,5,6$ give qualitatively similar results. $\phi_{b}^{(2)}$ is positive in the slate region and orange in the negative. $\phi_{b}^{(2)}$ is always positive for negative tension ETW branes.}
\end{figure}

We have analytically computed the smaller eigenvalue $\lambda_-$ in $d=2$ and numerically in $d=3,4,5,6$. In each of these $d > 2$ cases, the behavior is qualitatively the same as in the $d = 2$ computation; $\phi_{b}^{(2)} < 0$ in a subregion of parameter space for positive-tension ETW branes. We summarize our results in Fig.~\ref{LightCrossingChanges}. 

We also considered the case where the operator dual to $\varphi$ is irrelevant. In this case we must set the position-dependent source $\propto c_2=0$ and the perturbation $\phi_b^{(2)}$ is proportional to $c_1^2$. Analytically in $d=2$, and numerically in $d=3,4,5,6$ we found that the proportionality constant to always be positive. In other words, turning on the scalar profile always increases $\phi_b$.

\subsection{Linearized Janus}

Any CFT with a relevant scalar operator of dimension $\Delta$ admits Janus-like deformations in which we study the theory in flat space, single out one spatial direction $x_{\perp}$, and turn on a position-dependent source 
\begin{equation}
    \lambda(x) = \begin{cases} \frac{\lambda_+}{|x_{\perp}|^{d - \Delta}}\,, & x_{\perp}>0\,,
    \\
    \frac{\lambda_-}{|x_{\perp}|^{d - \Delta}}\,, & x_{\perp}<0\,,\end{cases}
\end{equation}
where $\lambda_{\pm}$ are constants. This profile is consistent with defect conformal symmetry with a defect at $x_{\perp} = 0$. The M-theory Janus background~\cite{DHoker:2009lky} is dual to such a deformation of the ABJM theory.

In this Appendix we study the gravity dual of a linearized version of this deformation in which $\lambda_{\pm}$ are small. We compute the perturbation of the light crossing time in this regime.

\begin{figure}[t]
    \centering
    \includegraphics[width=3.5in]{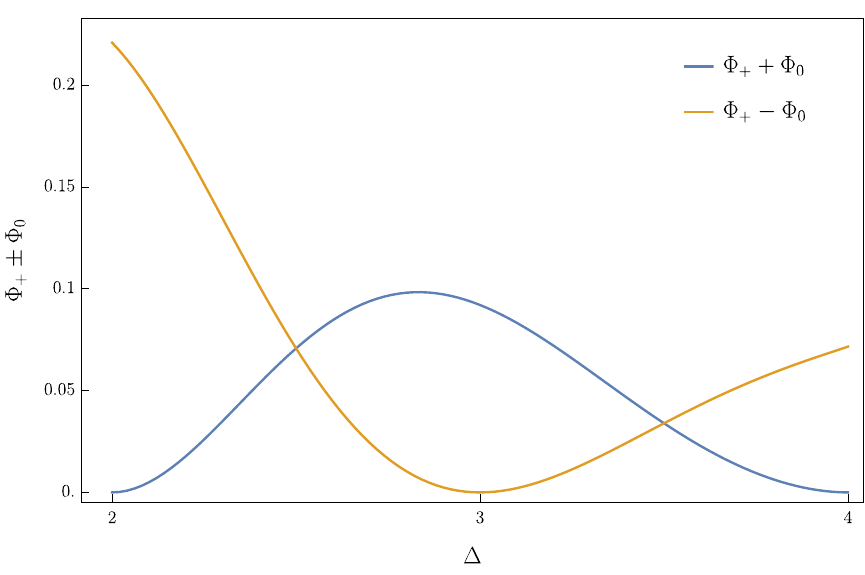}
    \caption{\label{LinearizedMassPlot} A numerical computation of the eigenvalues $\Phi_+\pm \Phi_0$ governing the perturbation in the light crossing time for linearized Janus deformations in $d=4$. }
\end{figure}

To compute the background and the light crossing time we rely heavily on the previous Appendix. The background is described by~\eqref{E:perturbExpansion} with the scalar described by~\eqref{E:phi1}, which we can parameterize instead as
\begin{equation}
    \varphi^{(1)}(\rho) = \text{sech}^{d/2}(\rho)\left( c_+ P_{\frac{d}{2}-1}^{\frac{d}{2}-\Delta}(\tanh(\rho))+ c_- P_{\frac{d}{2}-1}^{\frac{d}{2}-\Delta}(-\tanh(\rho))\right)\,,
\end{equation}
and which determines the perturbation in the warpfactor through~\eqref{E:newA2}. The perturbation in the light crossing time simplifies to
\begin{equation}\label{massJanusPhib}
    \phi_{b}^{(2)} = -\int_{-\infty}^{\infty} d\rho\,\tilde{S}_2(\rho)\,,
\end{equation}
with $\widetilde{S}_2(\rho)$ defined in~\eqref{E:regularizedS}. By virtue of a parity symmetry under $\rho \to -\rho$, we write this perturbation as
\begin{equation}
    \phi_b^{(2)} = \begin{pmatrix} c_+ & c_- \end{pmatrix} \begin{pmatrix} \Phi_+ & \Phi_0 \\ \Phi_0 & \Phi_+ \end{pmatrix}\begin{pmatrix} c_+ \\ c_- \end{pmatrix}\,,
\end{equation}
for complicated integrals $\Phi_+$ and $\Phi_0$. The eigenvalues $\Phi_+ \pm \Phi_0$ parameterize the change in the light crossing time. Numerically we find that these eigenvalues are always positive in the range $\frac{d}{2}\leq \Delta \leq d$, indicating that a linearized Janus deformation always increases the light crossing time from its pure AdS value of $\pi$. We plot the dependence of these eigenvalues with $\Delta$ for $d=4$ in Fig.~\ref{LinearizedMassPlot}.

\subsection{Massless scalar field}\label{MasslessFieldBridge}

In this Appendix we consider Einstein gravity with negative cosmological constant coupled to a massless scalar field dual to a marginal operator in a putative dual CFT. The action is
\begin{equation}
    S=\int d^{d+1}x \sqrt{-g}\left(R-d(d-1)-\frac{1}{2}(\partial \varphi)^2\right) \,,
\end{equation}
which has AdS$_{d+1}$ solutions with unit radius. This model can be embedded into string theory in $d=4$ where it leads to the non-SUSY Janus solution. Here we study its variants and their light crossing times in dimensions $d=2,3,4,5,6$.

We are interested in the dual of a Janus-like interface preserving defect conformal symmetry. The background has the form~\eqref{E:BCFTbackground}. The Klein-Gordon equation for $\varphi$ is solved by
\begin{equation}
    \varphi' = Q e^{-dA}\,,
\end{equation}
for constant $Q$ labeling the scalar ``flux.'' We solve the $\rho\rho$-component of the Einstein's equations
\begin{equation}
\label{E:scalarE1}
    d(d-1)\left(A'^2+e^{-2A}-1\right)=Q^2 e^{-2dA}\,,
\end{equation}
which implies the other components. In $d=2$ this equation can be solved analytically to give two possible solutions related to each other by $\rho\to - \rho$,
\begin{equation}
    e^{2A} =\cosh^2(\rho) - \frac{Q^2}{2}e^{\pm 2\rho}\,,
\end{equation}
while in general $d$ it is insoluble in terms of elementary functions. Since $e^{2A}>0$, we  have an upper bound on $Q^2$ and in $d=2$ reads $Q \leq \frac{1}{\sqrt{2}}$. In general $d$ this bound can be determined in the following way. Consider also the component of the Einstein's equations obtained by double contraction with the null vector $U^{\mu}\partial_{\mu} = z e^{-A}\partial_t + \partial_{\rho}$ (in coordinates where $ds_{\text{AdS}_d}^2 = \frac{-dt^2+dz^2 + d\vec{x}^2}{z^2}$),
\begin{equation}
\label{E:scalarE2}
    (1-d)(A''-e^{-2A}) = Q^2 e^{-2dA}\,.
\end{equation}
As $\rho \to \pm \infty$ we have $A(\rho) \sim \pm \rho$, in which case it has a minimum value at some $\rho = \rho_t$ with $A'(\rho_t) = 0$. At this point we should have $A''(\rho_t) \geq 0$ and $e^{2A(\rho_t)}\geq 0$. The $\rho\rho$-component~\eqref{E:scalarE1} implies
\begin{equation}
    Q^2 = d(d-1) \frac{x-1}{x^d}\,, \qquad x =e^{-2A(\rho_t)}\,,
\end{equation}
while using~\eqref{E:scalarE2} the condition that $\rho_t$ is a minimum of $A$, $A''(\rho_t)\geq 0$ implies
\begin{equation}
    A''(\rho_t) = d-x(d-1) \geq 0\,, \qquad \Rightarrow\qquad x \leq \frac{d}{d-1}\,,
\end{equation}
with $A''(\rho_t)=0$ when $x = \frac{d}{d-1}$. Using 
\begin{equation}
    \frac{dQ^2}{dx} = d(d-1)x^{-(d+1)}(d-x(d-1)) = d(d-1)x^{-(d+1)}A''(\rho_t) \,,
\end{equation}
we see that $Q^2$ grows monotonically with $x$, which tells us that the maximum value of $Q^2$ is attained at the largest possible value of $x$, $\frac{d}{d-1}$. This gives
\begin{equation}
    Q^2 \leq Q^2_{\rm max} = d\left( \frac{d-1}{d}\right)^d\,.
\end{equation}

\begin{figure}[tp]
    \centering
    \begin{minipage}{.45\textwidth}
    \includegraphics[width=2.7in]{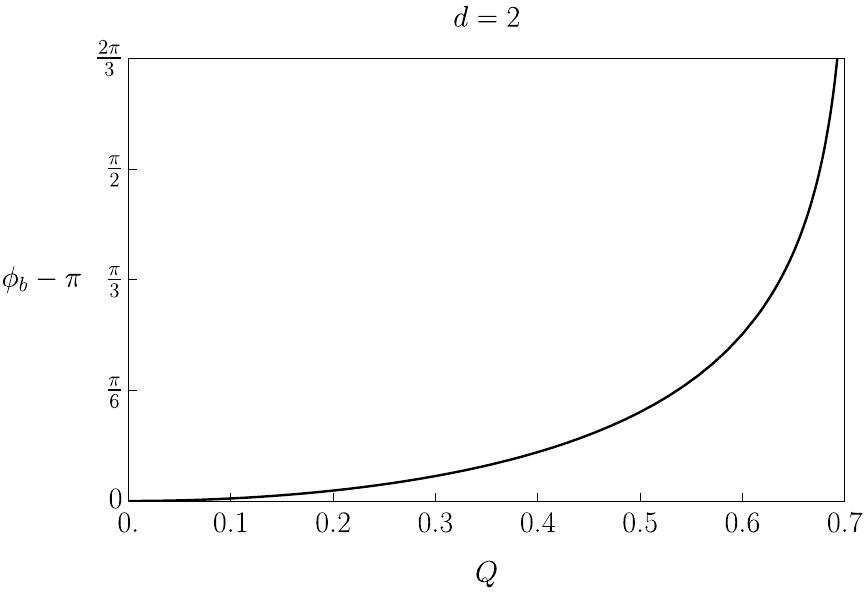}
    \end{minipage}
    \hfill
    \begin{minipage}{.45\textwidth}
    \includegraphics[width=2.7in]{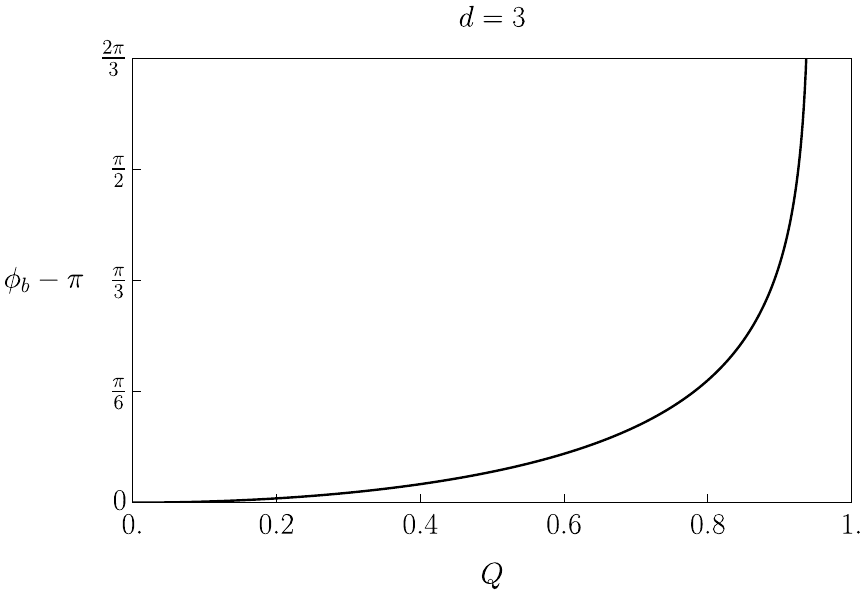}
    \end{minipage}
    \\
    \begin{minipage}{.45\textwidth}
    \includegraphics[width=2.7in]{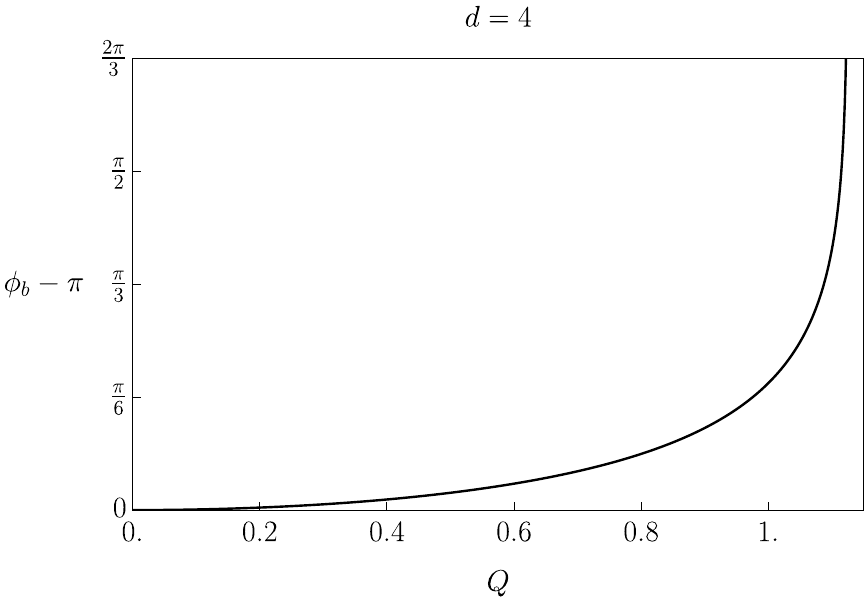}
    \end{minipage}
    \hfill
    \begin{minipage}{.45\textwidth}
    \includegraphics[width=2.7in]{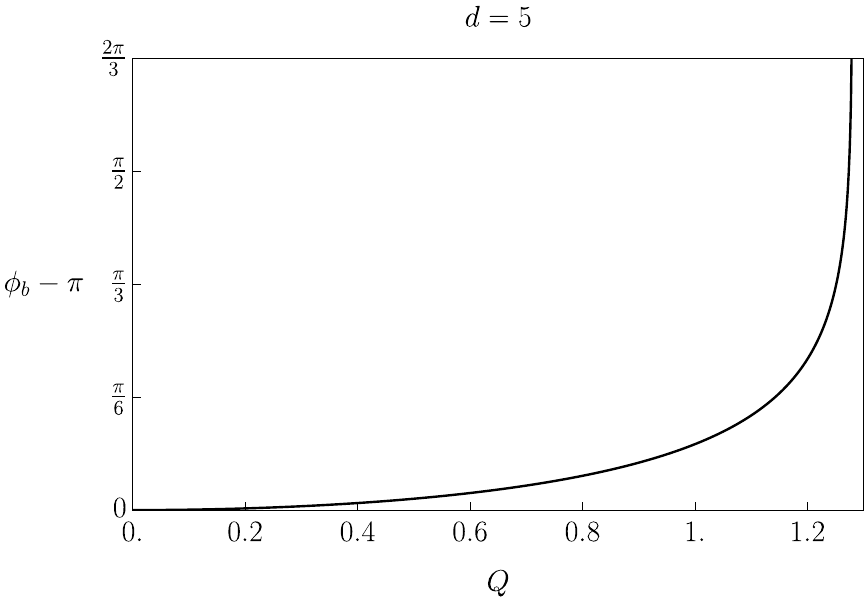}
    \end{minipage}
    \\
    \includegraphics[width=2.7in]{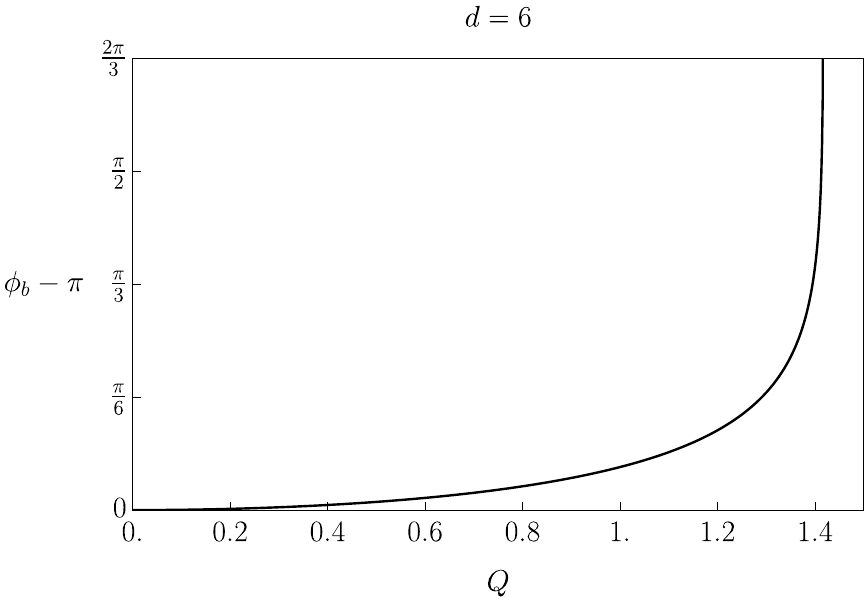}\,
    \caption{\label{NumbericalMassess} The light crossing time $\phi_b-\pi$ as a function of scalar flux $Q$. The $d = 2$ case is an analytical plot made from \eqref{2dJanussol}, while $d = 3, 4, 5, 6$ were computed numerically; we note our numerical methods are consistent with the analytical $d = 2$ solution.}
\end{figure}

With the warpfactor in hand we can compute the light crossing time. In $d=2$ we do so analytically with the result
\begin{equation}\label{2dJanussol}
 \phi_{b} = \frac{2  \sqrt{\sqrt{1-2 Q^2}+1}}{| Q| } \text{Re}\left[K\left(\frac{-2 Q^2+2 \sqrt{1-2 Q^2}+2}{2 Q^2}\right)\right]\,,
\end{equation}
where $K$ is the complete elliptic integral of the first kind. In $d=3,4,5,6$ we resort to numerics both to find the background and then to perform the light-crossing integral. In all cases we find $\phi_b \geq \pi$. See Fig.~\ref{NumbericalMassess}. 

\section{Top-down entropy}

In the main text we used results for the defect/boundary entropy of several top-down models. We summarize those results here.

The defect entropy of the non-SUSY Janus solution was computed in~\cite{Estes:2014hka} to be
\begin{equation}
    D_0 = N_3^2 \left(-\frac{1}{4}+(\zeta(\xi_0) -\sqrt{\gamma})\xi_0 - \frac{1}{2}\ln \left(\frac{\sigma(2\xi_0)}{2\sqrt{\gamma}} \right)+ \frac{\sqrt{\gamma}}{2}\zeta(2\xi_0)\right)\,.
\end{equation}
Here $\gamma \in [3/4,1]$ labels the solution; $\zeta(\xi)$ and $\sigma(\xi)$ are the Weierstrass $\zeta$ and $\sigma$ functions; and the non-SUSY Janus solution~\eqref{E:metricnonSUSYJanus} exists for $\xi \in (-\xi_0,\xi_0)$. 

The defect entropy of the D1/D5 SUSY Janus DCFT was computed in~\cite{Chiodaroli:2010ur} and is given by
\begin{equation}
    D_0 = \frac{c}{3}\ln \kappa\,, \qquad c = \frac{12\pi^2 L^2}{G_6}\,,
\end{equation}
where $c$ is the central charge of the dual CFT and $(L,\kappa)$ characterize the 6d geometry~\eqref{reducedD1D5}.

The defect entropy of the D3/D5 DCFT was computed in~\cite{Estes:2014hka} and is given by 
\begin{align}\label{D3D5IntEntropy}
\begin{split}
    D_0&=\frac{1}{4}\left[(N_3^+)^2\ln\left(\frac{g_{\rm YM}^2\alpha^2}{\pi^2N_3^+}\right) + (N_3^-)^2\ln\left(\frac{g_{\rm YM}^2\alpha^2}{\pi^2N_3^-}\right) \right]\\
    &+ \frac{g_{\rm YM}^4N_5\alpha^3}{12\pi^4}\left(\cosh(3\delta)-6\cosh(\delta)+12\delta\sinh(\delta)\right)\\
    &+ \frac{g_{\rm YM}^4N_5^2\alpha^2}{16\pi^4}\left(4\delta\sinh(2\delta)-3\cosh(2\delta)+8\ln{2}\sinh^2(\delta)\right) \,,
\end{split}
\end{align}
with $(\alpha,\delta)$ determined by field theory parameters through~\eqref{D3D5parameters}. In the main text we studied the limit $\delta = 0$ in which case the 5d cosmological constant does not change across the defect. In that case~\eqref{D3D5IntEntropy} becomes
\begin{equation}
    D_0=\frac{N_3^2}{2}\ln\left(\frac{g_{\rm YM}^2\alpha^2}{\pi^2N_3}\right) - \frac{5g_{\rm YM}^4N_5\alpha^3}{12\pi^4}- \frac{3g_{\rm YM}^4N_5^2\alpha^2}{16\pi^4} \,.
\end{equation}
The D3/D5 BCFT is also described by the $N_3^-\to 0$ limit of the D3/D5 defect entropy. The result described in~\cite{Estes:2014hka} is
\begin{equation}
\label{E:D3D5B}
    B_0 = - \frac{N_3^2}{8}\left( 2 \ln \left( \frac{8 \underline{\alpha}}{N_5}\right) -3\right) - \frac{\underline{\alpha}}{3N_5}\,,
\end{equation}
where $\underline{\alpha}$ is determined by field theory parameters through~\eqref{E:hattedAlphas}.

The defect entropy of the SUSY Janus interface was computed in~\cite{Estes:2014hka} to be
\begin{equation}
    D_0 = \frac{N_3^2}{2}\ln (\cosh(\delta\phi))\,.
\end{equation}

\bibliography{refs}
\bibliographystyle{JHEP}
\end{document}